\documentstyle[12pt,draft,epsf]{article}

\newcommand \be  {\begin{equation}}
\newcommand \ee  {\end{equation}}
\newcommand \bea {\begin{eqnarray} \nonumber }
\newcommand \eea {\end{eqnarray}}

\newcommand \al  {\alpha}
\newcommand \de  {\delta}
\newcommand \eps {\epsilon}

\newcommand \la  {\lambda}
\newcommand \La  {\Lambda}
\newcommand \s   {\sigma}

\newcommand \cD  {{\cal D}}
\newcommand \cH  {{\cal H}}
\newcommand \cN  {{\cal N}}

\newcommand \tC  { {\tilde{C}} }
\newcommand \ba  { \overline }

\newcommand \dbbb { {dB d\overline{B}} }
\newcommand \dccc { {dC d\overline{C}} }

\begin{document}

\title{Replica Field Theory for Deterministic Models: Binary
       Sequences with Low Autocorrelation}
\author{Enzo Marinari$^{(a,b)}$, Giorgio Parisi$^{(c)}$
        and Felix Ritort$^{(a,c)}$\\[0.5em]
  {\small (a): Dipartimento di Fisica and Infn, Universit\`a di Roma
    {\em Tor Vergata}}\\
  {\small \ \  Viale della Ricerca Scientifica, 00133 Roma (Italy)}\\
  {\small (b): NPAC, Syracuse University, Syracuse NY 13210 (USA)}\\
  {\small (c): Dipartimento di Fisica and Infn, Universit\`a di Roma
    {\em La Sapienza}}\\
  {\small \ \  P. A. Moro 2, 00187 Roma (Italy)}\\[0.5em]}
\date{May 20, 1994}
\maketitle

\begin{abstract}

  We study systems without quenched disorder with a complex landscape,
and we use replica symmetry theory to describe them. We discuss the
Golay-Bernasconi-Derrida approximation of the low autocorrelation
model, and we reconstruct it by using replica calculations. Then we
consider the full model, its low $T$ properties (with the help of
number theory) and a Hartree-Fock resummation of the high-temperature
series. We show that replica theory allows to solve the model in the
high $T$ phase. Our solution is based on one-link integral techniques,
and is based on substituting a Fourier transform with a generic
unitary transformation. We discuss this approach as a powerful tool to
describe systems with a complex landscape in the absence of quenched
disorder.

\end{abstract}
\vfill
\begin{flushright}
  {\bf  hep-th/9405148}\\
  {\bf  \hfill ROM2F/94/015}\\
  {\bf  \hfill Roma-La Sapienza 1019}\\
\end{flushright}
\newpage

%%%%%%%%%%%%%%%%%%%%%%%%%%%%%%%%%%%%%%%%%%%%%%%%%%%%%%%%%%%%%%%%%%%%%%%%%%
\section {Introduction\protect\label{S_INT}}

This note has been prompted by two main motivations. One comes from a
problem whose solution has relevant practical applications, while the
other one is more abstract in nature, and is generated from what we
have learned in the last years about disordered
systems~\cite{LIBRO1,LIBRO2}.

We will be dealing with the problem of finding binary sequence with
low autocorrelation~\cite{LOWOLD,GOLAY,BERNAS}. Sequences of this kind
are important in favoring efficient communication, and the practical
side of the problem is obvious. We hope we will convince the reader it
is also fascinating from a theoretic point of view.

When we search binary sequences of $+1$ and $-1$ having minimal
autocorrelation we are dealing with a completely deterministic
problem, and disorder is not a part of the game. In our starting rules
there is nothing random. Still, we will see how the system can indeed
have a behavior that is very much reminiscent of a random
system. Changing one spin to optimize a given set of correlations can
increase other correlation functions, with a competitive effect which
turns out to be typical of a system which contains disordered
couplings. We will see that replica symmetry
theory~\cite{LIBRO1,LIBRO2} can be an useful tool even for describing
this kind of systems. We will be able, by using the analogy with a
relevant disordered system, to capture the general features of the
model. We will try to understand and stress the differences which
distinguish a low autocorrelation model from a spin glass like
model. That will lead us to a detailed discussion of the low
temperature properties of the low autocorrelation model.

We present a careful investigation of some statistical mechanics
aspects of the problem, by largely extending previous results due to
Golay~\cite{GOLAY} and to Bernasconi~\cite{BERNAS}. We establish a
relation between this deterministic problem and random spin glasses,
which we consider a very interesting outcome of this study. Some ideas
typical of spin glasses, as replica symmetry breaking, can be
successfully used in this context.

In section (\ref{S_DEF}) we define the models we will discuss in the
rest of the paper. In section (\ref{S_GRO}) we discuss the ground
state structure of the model (also by using well known number theory,
see for example \cite{PRIME}) and we begin a discussion of its phase
diagram and of the low temperature phase. In section (\ref{S_APP}) we
discuss the validity of the Golay-Bernasconi approximation. We
introduce the replica symmetry approach, we define a disordered model
and we study its behavior. In section (\ref{S_HIG}) we investigate in
better detail the high-temperature regime. We perform and describe a
high-temperature expansion. We introduce a Hartree-Fock approximation
which allows us to write a closed form for the free energy.

In section (\ref{S_ALL}) we discuss the full phase diagram of the model. In
section (\ref{S_REP}) we introduce one more model which can be solved by using
the replica approach. The solution is the same we get with the Hartree-Fock
approximation.  In section (\ref{S_CON}) we draw our conclusions.

The reader which will find this problem interesting will be happy to
know that much related material is becoming available. Reference
\cite{MIGLIO} mainly contains a study of the dynamical properties of
the system, which uses the {\em tempering} Monte Carlo
approach~\cite{TEMPER}. Ref. \cite{MIGRIT} discusses aging in low
autocorrelation models. Reference \cite{MAPARI1,MAPARI2} introduces and
discusses more models and analogies with random systems (and, in
particular, the open low autocorrelation model, see later). More
results, which partially overlap with ours, will be discussed by
Bouchaud and Mezard in \cite{BOUMEZ}.

%%%%%%%%%%%%%%%%%%%%%%%%%%%%%%%%%%%%%%%%%%%%%%%%%%%%%%%%%%%%%%%%%%%%%%%%%%
\section{Definition of the Model\protect\label{S_DEF}}

Let us consider a sequence of length $N$ of spin variables $\s_j$.  They are
labeled by a one-dimensional index $j$ ($\s_j$, $j=1,N$), and can take the
values $\pm 1$. The Hamiltonian is defined by

\be
  \protect\label{E_H1}
  H \equiv \frac{1}{N-1} \sum_{k=1}^{N-1} C^2_k\ ,
\ee

\noindent where $C_k$ is the sum of the $\s_i-\s_j$ correlation functions at
distance $k\equiv |i-j|$. The choice of the boundary conditions, i.e. of
the terms we will include in the sum (\ref{E_H1}), allows us to define two
different models.

\begin{itemize}
\item
The {\bf open} model is defined by using open boundary conditions. In
this case $C_k$ is obtained by summing $N-k$ terms:

\be
  \protect\label{E_COPEN}
  C_k \equiv \sum_{j=1}^{N-k} \s_j \s_{j+k}\ .
\ee

\item
The {\bf periodic} model is defined by using periodic boundary
conditions. Here we are considering a closed chain, and:

\be
  \protect\label{E_CPERIODIC}
  C_k \equiv \sum_{j=1}^{N} \s_j \s_{(j+k-1)(\mbox{\footnotesize mod} N)+1}\ .
\ee

Here we have summed $N$ contributions, considering all spin couples at
distance $k$ on the closed chain.
\end{itemize}

The periodic model has some peculiarities which allow us to study it
in greater detail. The main tool we will use is the Fourier
transform. We can rewrite the periodic Hamiltonian as

\be
  \protect\label{E_HF1}
  H = {1 \over N-1}\sum_{p=1}^{N} (|B(p)|^4 -1) +1\ ,
\ee

\noindent where the $B(p)$ are the Fourier transformed $\s_i$, and the
Fourier transform is defined as

\be
  \protect\label{E_FT}
  B(p) \equiv \frac{1}{\sqrt{N}}
  \sum_{j=1}^{N} e^{i \frac{2 \pi p}{N} j} \s_j\ .
\ee

\noindent
In eq. (\ref{E_HF1}) we had to subtract a constant factor since in the
sum of eq. (\ref{E_H1}) we do not include the constant correlation at
distance zero.

In this paper we will focus on the periodic model. Further results about the
open model will be contained in ref. \cite{MAPARI2}.

As we have already discussed much attention has been devoted in the past to
the problem of finding the ground state of such a model
\cite{LOWOLD,GOLAY,BERNAS}. Here we will continue such an effort, but we will
also (and mainly) extend our study to the thermo-dynamical behavior of
the model.  We will study its behavior as a function of the inverse
temperature $\beta \equiv \frac{1}{T}$. Our main efforts will be
devoted to the computation of free energy density. We define the
partition function of our system as

\be
  \protect\label{E_ZETA}
  Z_N(\beta) \equiv  \sum_{\{\s\}} e^{-\beta H(\{\s\})}\ ,
\ee

\noindent where the sum runs over the $2^N$ allowed configurations of the spin
variables, and the free energy density as

\be
  \protect\label{E_FREE}
  f(\beta)  \equiv \lim_{N\to \infty}\ ( - {1 \over \beta N}
    \ \ln(Z_N(\beta))  ) \ .
\ee

Once again, we note that this approach has both a practical interest
and a theoretical one. It is interesting to study the full
thermo-dynamical behavior of the system since that gives more
information about features of the low autocorrelation sequences. We
will be interested for example in their number and their basin of
attraction, and in their stability properties (which can be very
relevant for practical applications). On the other side such a
statistical mechanics approach will help us to shift towards the realm
of disordered systems.

%%%%%%%%%%%%%%%%%%%%%%%%%%%%%%%%%%%%%%%%%%%%%%%%%%%%%%%%%%%%%%%%%%%%%%%%%%
\section{The Ground State Energy and a First Look at
Thermodynamics\protect\label{S_GRO}}

The ground state of the periodic model defined by the Hamiltonian
(\ref{E_H1}) (with $C_k$ given by (\ref{E_CPERIODIC})) is not known in
general. No systematic procedure to construct ground state
configurations for generic $N$ is known. A remarkable exception holds
for given values of $N$, where {\em ad hoc} constructions exist. Such
constructions are mainly based on number theory
\cite{PRIME}, and they produce spin sequences with a total energy of order $1$,
i.e. with an energy density $e \equiv \frac{H}{N}$ of order
$\frac{1}{N}$ (which tends to zero in the thermo-dynamical limit).

Let us describe a simple construction\footnote{The same spin sequence
can be obtained by using directly Legendre quadratic residues
\cite{PRIME}. For all positive integer $j<N$ we compute $J \equiv
(j\cdot j)$ (mod $N$), and we set $\s_J=+1$. In all locations but the
$N$-th one (where we set $\s_N=0$) that cannot be obtained through
this procedure we set $\s_I=-1$.}, which works when $N$ is a prime
larger than $2$ \cite{PRIME}. We set the $\s_j$ variables to $-1$, $0$
or $+1$ by identifying

\be
  \s_j= j^{{1 \over 2}(N-1)}\mbox{\rm mod } N\ .
\ee

\noindent
In this way we get\footnote {A theorem by Fermat \cite{PRIME} tells us
that if $j$ is not a multiple of $N$ than $j^{(N-1)}=1$, mod
$N$. Therefore in this case $j^{{1 \over 2}(N-1)} =
\pm 1$.} $\s_j=\pm 1$ for $j<N$, and $\s_N=0$.  For example for $N=13$
by using this construction we get the sequence

$$
  \protect\label{E_N13} \nonumber
  \begin{array}{cccccccccccccc}
   j    & 1  & 2  &  3 &  4 &  5 &  6 &  7 &  8 &  9 & 10 & 11 & 12 & 13    \\
   \s_j & +1 & -1 & +1 & +1 & -1 & -1 & -1 & -1 & +1 & +1 & -1 & +1 &  0
  \end{array}
$$

By following this procedure we have obtained a sequence which, but for
its last spin, is a legitimate one (in the sense it is composed by
$\pm 1$). Now we will proceed by first evaluating the energy of this
quasi-legal sequence, and eventually by computing the effect of
modifying the last spin to $\pm 1$, to get a truly legal sequence. We
will show that such a sequence is in some cases a true ground state
(i.e. it has the minimum allowed energy).

Computing the energy of such a sequence is an easy task.  Theorems
well known by mathematicians \cite{PRIME} tell us that in this case
all correlation functions $C_k$ are equal to $-1$ (we remind the
reader we are discussing the periodic model). We can also use a Gauss
theorem \cite{PRIME} to notice that the Fourier transformed variables
take here the form

\be
  B(p) = G(N)\  \s_p\ ,
\ee

\noindent
where $G(N)=1$ if the prime $N$ has the form $4n+1$ (with positive
integer $n$), and $G(N)=-i$ if it has the form $4n+3$ (in different
words on our sequences the Fourier transformed variables are equal or
proportional to the original $x$-space variables). It is clear that
the Hamiltonian (\ref{E_H1}) of the periodic model takes on our
slightly-illegal spin sequence the value $1$.

Now we have to understand what happens when we modify the spin $\s_N$,
by setting it to $\pm 1$. It is easy to see that when we do that the
Hamiltonian changes of a finite amount. Indeed for $N$ of the form
$4n+3$ the Hamiltonian does not change, and keeps it value of $1$. The
point is that (as can be easily verified by inspection) the $\pm 1$
sequences are in this case antisymmetric around the site $N$. For $N$
of the form $4n+1$ the $\pm 1$ sequences are symmetric around the site
$N$, and on the fully-legal sequence $H$ takes a value of $5$.

Since we are considering $N$ odd, it is clear that for $N$ prime of
the form $4n+3$ the two fully legal sequences we have built (and the
sequences obtained by using the translational invariance of the
problem, and the $\pm 1$ symmetry) are true ground states. This is
because for $N$ odd the minimum value allowed for each $C_k$ is $1$,
and the minimum value allowed for $H$ is $1$. We have exhibited
configurations with the minimal allowed energy, i.e. ground states.

Let us state again our conclusion. In the case of $N$ prime of the
form $4n+3$ we have obtained a thermodynamical ground state, whose
energy density goes to zero when the volume goes to
infinity. Translational invariance and spin flip invariance imply that
the degeneracy of the ground state is at least $2N$.

For other values of $N$, for example of the form $N=2^p-1$, there are
alternative techniques to construct the ground state, based for
example on the theory of Galois fields \cite{PRIME}. For example for
$N=2^{57}-1 = 144115188075855871$ one finds that the sequence which
satisfies the relation

\be
  \s_j=\s_{j-24}\s_{j-57}
\ee

\noindent is a ground state. If we exclude the trivial case of
$\s_i$ identically equal to $1$ (which is not a ground state), such
sequence is unique, apart from a translation\footnote{The sequence is
specified by its first $p=57$ elements.  Therefore there are $2^p-1$
different sequences, which is exactly the number of possible
translations. It can be shown that every subsequence of $p$ elements
appears once and only once, apart from the subsequence with all $1$,
which is forbidden.}~\cite{PRIME,KNUTH}.

It is rather interesting to note that also in this case the Fourier
transform is very similar to the original sequence. One finds that it
exists a value of $s$ such that

\be
  B(p)=\s_{p+s}\ .
\ee

\noindent
The deep reasons for this duality among configuration and Fourier
space escape us.

It is quite remarkable that this last sequence is considered at all
practical effects a good random sequence (see for example
\cite{KNUTH}). We can summarize the status of things by saying that
the ground state of our model can be obtained as the {\em output of a
random number generator}!  This is surprising, but maybe not so
much. When designing a random number generator one wants bit sequences
with low autocorrelation. That means that for large values of $N$
the correlation functions should not be proportional to $N$. A true
sequence of random numbers should have autocorrelations of order
$N^{\frac{1}{2}}$. One is doing ``better'' than that by obtaining
sequences with autocorrelation of order $1$. That does not seem to
cause any practical problem.

For generic values of $N$ we do not have any method to explicitly
exhibit the ground state, and we do not know the ground state energy.
The very existence of the thermodynamic limit is non trivial. One
could get different results when $N$ goes to infinity depending on the
arithmetic properties of $N$ sequence one selects.  We shall see later
that in the high-temperature region the $N^{-1}$ corrections are
different for sequences consisting of even or odd values of $N$.  The
corrective terms proportional to $N^{-3}$ also change depending if one
selects an $N$ series such that $N$ is or not multiple of $3$. We will
see that in general things become more and more complex when we look
at higher order corrections.

In order to get the first hints about the ground states and the
thermodynamical behavior of the system we have used two approaches. In
first we have solved exactly (by computing the density of states by
exact enumeration) systems of size up to $N=38$. By examining all
configurations we have computed the number of configurations of a
given energy $\cN (E)$ as a function of $E$. We have looked at the
ground state energy $E_0$, and stored and analyzed the ground state
and the first excited state configurations (at least for some of the
$N$ values).  From $\cN (E)$ we are able to reconstruct the partition
function, the free energy density and all the related thermodynamical
quantities.

As a second step we have looked for the ground state energy by using a
minimization procedure. For a given $N$ value we start from a random
$\s_i$ configuration, and we minimize its energy by single
spin-flips. We repeat this procedure until satisfaction. We assume we
have reached the ground state when the minimum energy has been found
$F$ times\footnote{We are using in \cite{LAMAPA} the same procedure to
try to find all solutions of the mean field equations for the Random
Field Ising Model in $3d$.}. In the case where we also have the exact
solution ($N \le 38$) this procedure easily gives the correct ground
state energy. The choice of $F=100$ recognitions is still safe in the
$N$ region going up to $N=50$. Low energy states with a small basin of
attraction are the most dangerous. For the case of the good prime
$N=47$ (where by good we mean here of the form $4n+3$) the first
excited state is found a number of times of order of $50$ before
finding the true ground state (which in this case, as we have
explained, we know exactly).

\begin{figure}
  \epsfxsize=400pt\epsffile{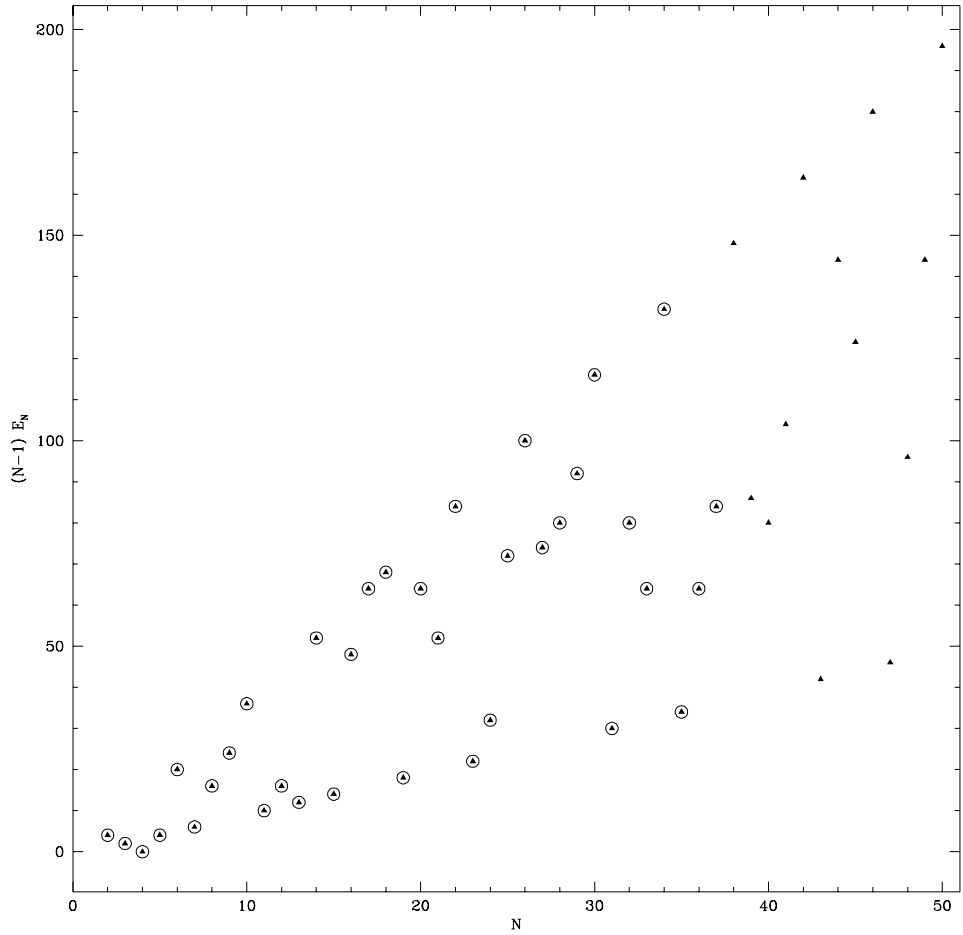}
  \caption[a]{\protect\label{F_E0N}
    The ground state energy times $(N-1)$ as a function of $N$. The
    small filled triangle are from the minimization search.
    For $N\le 38$ where we also have the exact solution
    the small triangles are circled by larger empty dots.
  }
\end{figure}

In fig. \ref{F_E0N} we plot ($N-1$) times the ground state energy as a
function of $N$. The small filled triangles are from the minimization
search. For $N\le 38$ they are circled by larger empty dots (that
reminds the reader that in this case we also have the exact result,
which coincides with the the minimization result).

At a first look the ground state energy ($E_0$) depends quite
randomly on $N$. But we notice some regular patterns which can be
of some importance.

\begin{itemize}

  \item

For $N$ prime of the form $4n+3$ the ground state energy is the one
given by the exact construction we have described before. This is a
test of our programs and procedures.

  \item

For $N$ of the form $4n+2$, $n$ zero and positive integer, i.e. for
all the $n$ we have analyzed, we find

\be
  \protect\label{E_E4NP2}
  E_N = 4\ .
\ee

\noindent
We cannot be sure that this behavior is not an accident, but we have
to notice we find it for all values of $N$ of this kind.

\item

For $N$ of the form $4n+1$, $n\ge 8$, i.e. for $N\ge 33$, we have found that

\be
  \protect\label{E_E4NP1}
  E_N = 5 - \frac{96}{N-1}\ .
\ee

\noindent
For $N$ of this form, even for $N$ prime, our number theory based
ground state construction does not necessarily give a ground state.

\end{itemize}

\noindent
We can use these results to try some claims about the $N\to \infty$
limit for the ground state energy. The {\em merit factor}, used for
estimating how good a low autocorrelation sequence is, for a sequence
of length $N$ (and $N$ large, or to agree with standard definitions we
need to multiply times $N$ and divide times $N-1$) is given by

\be
  F^{(N)} \equiv \frac{N}{E^{(N)}}\ .
\ee

\noindent
If the energy goes to a constant value in the large $N$ limit that
means that the system will have a zero energy density, and a diverging
merit factor. We know that on the primes $N$ of the form $4n+3$ this
is exactly what happens. But we also know that such $N$ values have
zero measure, and selecting such a sequence could not be a reliable
way to go to the infinite volume limit for generic values of $N$.  If
the behavior we have described in equations
(\ref{E_E4NP2}),(\ref{E_E4NP1}) survives in the large $N$ limit we
have two finite measure sequences (including one $N$ value over $2$)
which have asymptotically a zero energy density. On the other $N$
values we are not able to draw even tentative and qualitative
conclusions like the above.

\begin{figure}
  \epsfxsize=400pt\epsffile{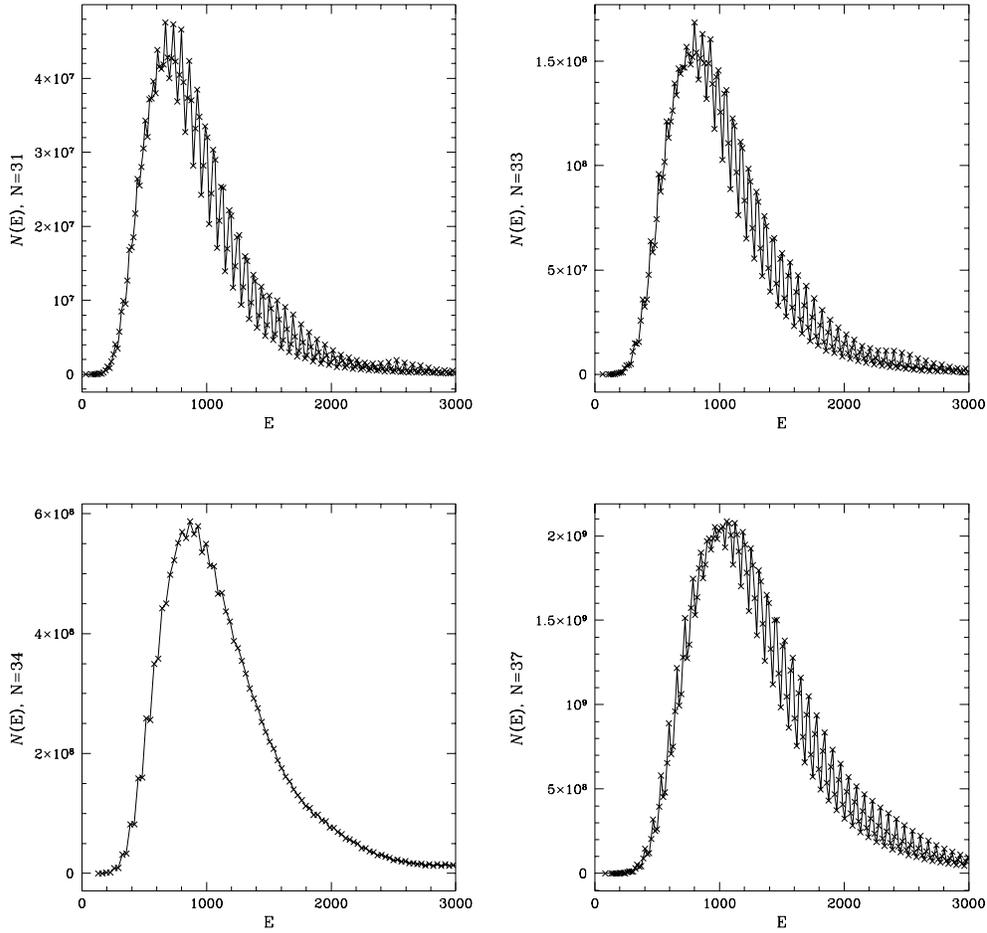}
  \caption[a]{\protect\label{F_NE}
    $\cN (E)$, the number of configuration of energy $E$ as a
    function of $E$. (a): $N=31$ (a good prime); (b) $N=33$ (of the
    form $4n+1$, non prime); (c): $N=34$ (of the form $4n+2$); (d): $N=37$
    (of the form $4n+1$, prime)
  }
\end{figure}

\begin{figure}
  \epsfxsize=400pt\epsffile{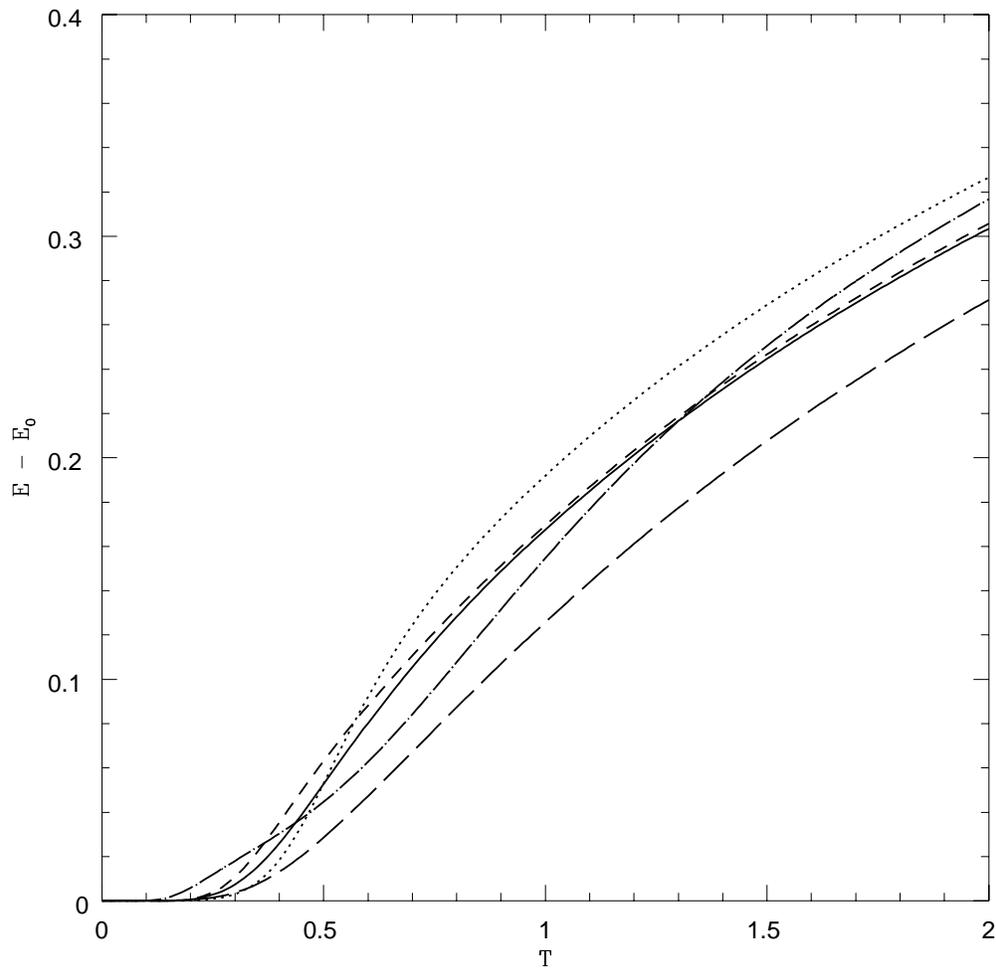}
  \caption[a]{\protect\label{F_ENET}
    The internal energy $E(T)$  minus the ground state energy
    (and normalized between zero and one) as a
    function of $T$, respectively for $N=19$ and $N=31$ (good primes,
    respectively dots-dashes and dots),
    $33$ (of the form
    $4n+1$, non prime, short dashes),
    $34$ (of the form $4n+2$, long dashes) and
    $37$ (of the form $4n+1$, prime, continuous line)
  }
\end{figure}

\begin{figure}
  \epsfxsize=400pt\epsffile{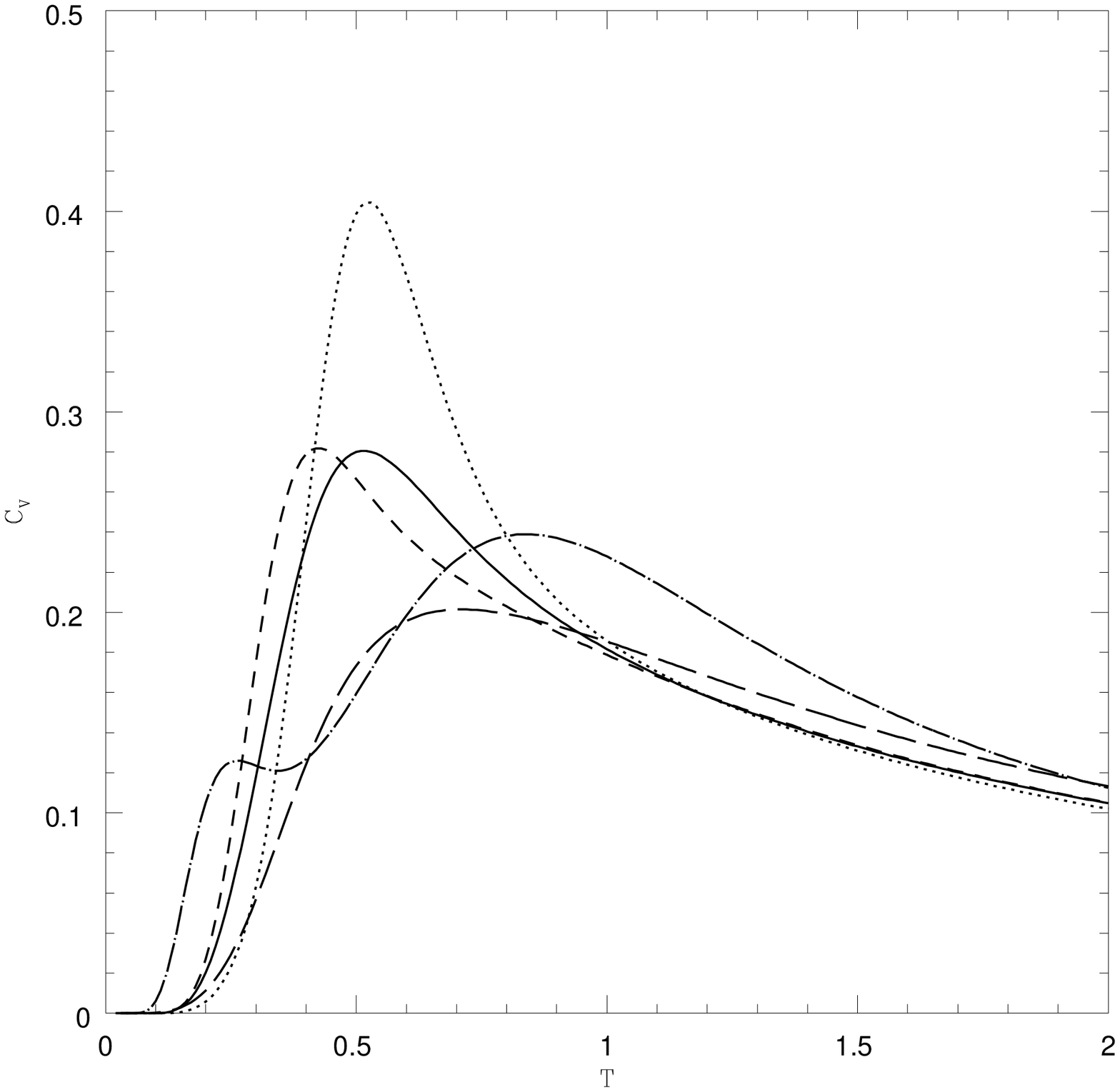}
  \caption[a]{\protect\label{F_CVT}
    As in figure (\ref{F_ENET}), but for the
    specific heat as a function of $T$.
  }
\end{figure}

The number of configurations of a given energy $\cN (E)$ allows us, as
we have explained, to evaluate the thermodynamical properties of the
system. In figures (\ref{F_NE}a-d) we show $\cN (E)$, the number of
configuration of energy $E$ as a function of $E$, respectively for
$N=31$ (a good prime), $33$ (of the form $4n+1$, non prime), $34$ (of
the form $4n+2$) and $37$ (of the form $4n+1$, prime). In figures
(\ref{F_ENET}) and (\ref{F_CVT}) we show respectively the internal
energy minus the ground state energy (normalized between zero and one)
and the specific heat as a function of $T$, for the same $N$
values and a smaller volume, $N=19$.

At this point we are able to draw a few tentative conclusions.

\begin{itemize}

\item

Changing $N$ of a small amount, typically of a $\Delta N = 1$ (when
$N$ is already of order $40$), induces large variations in
thermodynamic observable quantities in the low $T$ region.
Fluctuations from one volume size $N$ to a similar one are large, and
macroscopic. Such fluctuations forbid any simple extrapolation to the
$N\to\infty$ infinite volume limit. They decrease however for
increasing $N$. Their amplitude is compatible with being proportional
to $N^{-1}$ also at finite temperature.

\item

A pronounced peak in the specific heat increases with $N$, strongly
suggesting that in the infinite volume limit the system undergoes a
phase transition. The $T$ position of the maximum of the specific heat
decreases with increasing $N$ (in an irregular pattern). In the region
of $N\simeq 30-40$ from the position of the peak we estimate a
critical temperature $T_c \simeq 0.5$. The nature and the order of the
phase transition are difficult to assess.

\item

The density of states $\cN (E)$ for low energies depends on $E$
approximately as

\be
 \cN (E) = 2 N\  e^{A \ E}
\ee

\noindent (remember that the minimal degeneracy of the ground state is
$2N$). In our $N$ region $A$ turns out to be strongly dependent on
$N$. Such a dependence can be fitted well by a linear behavior.  This
is the same effect we can see in the $N$ dependence of the location of
the peak in fig. (\ref{F_CVT}).  For our large $N$ values (of order
$30-40$) the constant $A$ is of the order of $1.5$.

\item

The configurations with energy slightly larger that the ground state
energy are in average not similar to the ground state. The typical mutual
overlap of a ground state and a first excited state is not large when
$N$ increases\footnote{We define the overlap $q$ of two
configurations $\s$ and $\tau$ as $q \equiv\frac{1}{N} \sum_k \s_k
\tau_k$.}. In particular typical first excited state configurations
are not obtained by a single spin-flip operation on one of the ground
states. The configurations which are generated by a single spin-flip
on the ground state have in average energy higher than the first
excited state.  For example in the case of $N$ prime of the form
$4n+3$ the energy gap among the ground states and its one spin-flipped
excitation is at least of $3$. In this case no first excited state is a
single spin-flip of the ground state.

Let us analyze this point in better detail. For a ground state
configuration $s_0^\alpha$ (the series of the $N$ spin variables
$\sigma$ which form the ground state $\alpha$) we define the overlap
with the first excited state as

\be
  O^\alpha_{(0,1)} \equiv \frac{1}{N-2} \max_{A}
    (s_0^\alpha\cdot s_1^A)\ ,
\ee

\noindent
where $A$ runs over all first excited state configurations, $\alpha$
can take values over all ground state configurations, and the
$\cdot$ is the sum over sites of the product of the two spin
variables. $O^\alpha_{(0,1)}$ is $1$ when the ground state $\alpha$
corresponds to a first excited state which differs from the
configuration $\alpha$ in a single spin flip. This is the maximum
possible overlap. If there is the same number of equal spins and
different spins  $O^\alpha_{(0,1)}=0$. For a given $N$ value we define
the maximum overlap of the ground state and the first excited state as

\be
  O_{(0,1)}^M = \max_{\alpha}  O^\alpha_{(0,1)} \ ,
\ee

\noindent
where the maximum is taken over all configurations which have the
minimum energy. We plot $O_{(0,1)}^M$ as a function of $N$ in fig.
(\ref{F_MAX_OVER}). The maximum overlap is $1$ only for a few values
of $N$ (for large $N$, the ones of the form $4n+2$). For good primes
it is always very low.

\begin{figure}
  \epsfxsize=400pt\epsffile{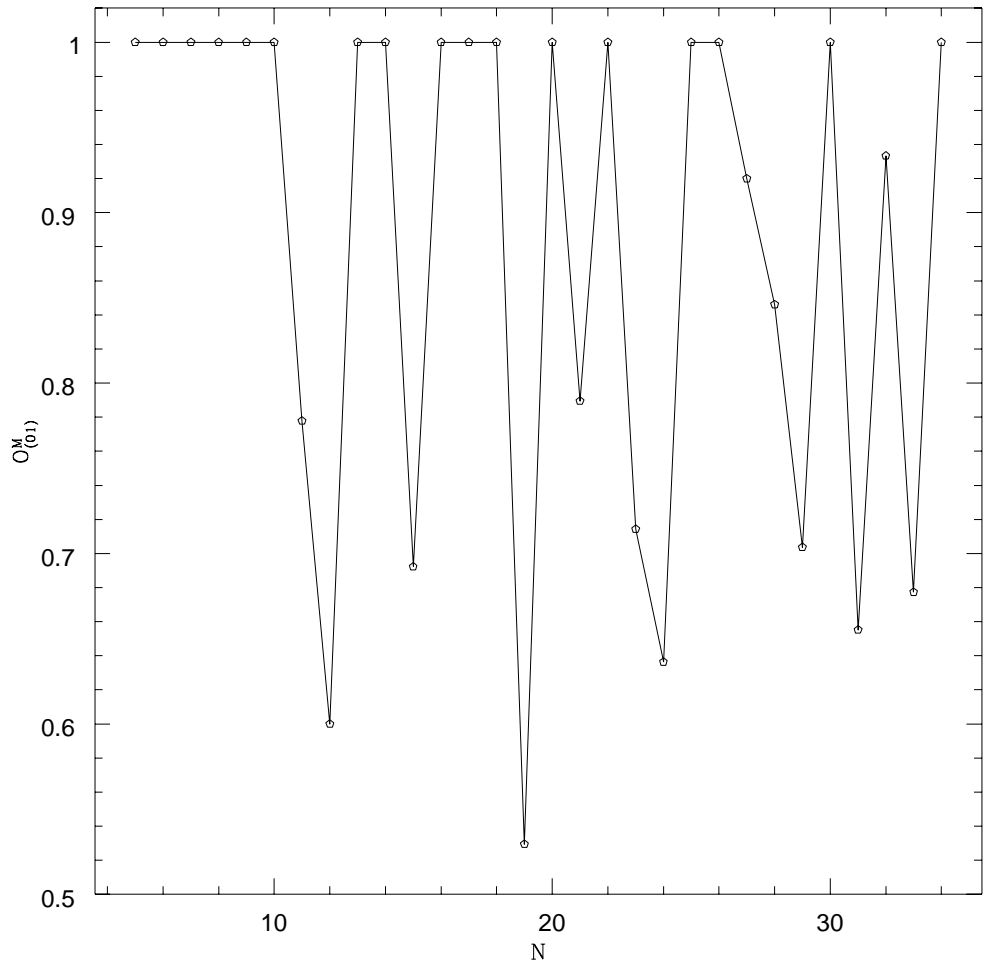}
  \caption[a]{\protect\label{F_MAX_OVER}
  $O_{(0,1)}^M$ as a function of $N$. }
\end{figure}

A more useful information can be gathered if we look at the average
ground state to first excited state overlap. We define

\be
  \langle O_{(0,1)}\rangle \equiv \frac{1}{\cN_0}
    \sum_\alpha O^\alpha_{(0,1)}\ ,
\ee

where $\cN_0$ is the sum runs over all ground states and $\cN_0$ is
their number. We plot $\langle O_{(0,1)}\rangle$ in
fig. (\ref{F_AVE_OVER}). As $N$ increases the average overlap
decreases, and for $N>22$ we never find a very large average overlap
between ground and first excited states.

\begin{figure}
  \epsfxsize=400pt\epsffile{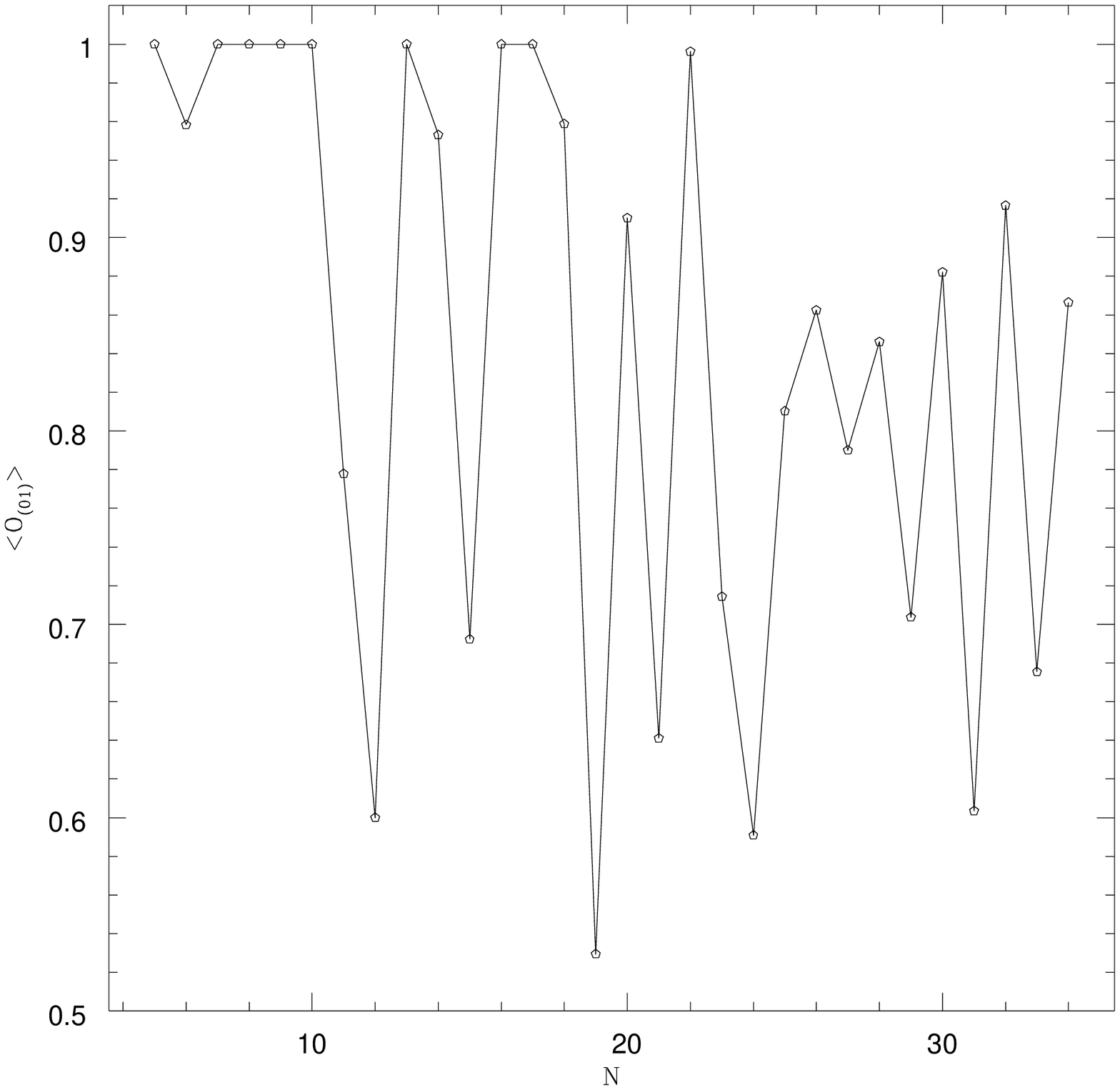}
  \caption[a]{\protect\label{F_AVE_OVER}
  $ \langle O_{(0,1)}\rangle$ as a function of $N$. }
\end{figure}

At last we plot in fig. (\ref{F_SF_ENE}) the ground state energies,
the first excited state energies and the average energy of
configurations obtained by a single spin flip from the ground state
(all of them multiplied times $(N-1)$).  The difference between single
spin flip and first excited state is large, and in this case (even
more than in fig. (\ref{F_AVE_OVER})) the effect does not depend
dramatically from the cardinality of $N$.

\begin{figure}
  \epsfxsize=400pt\epsffile{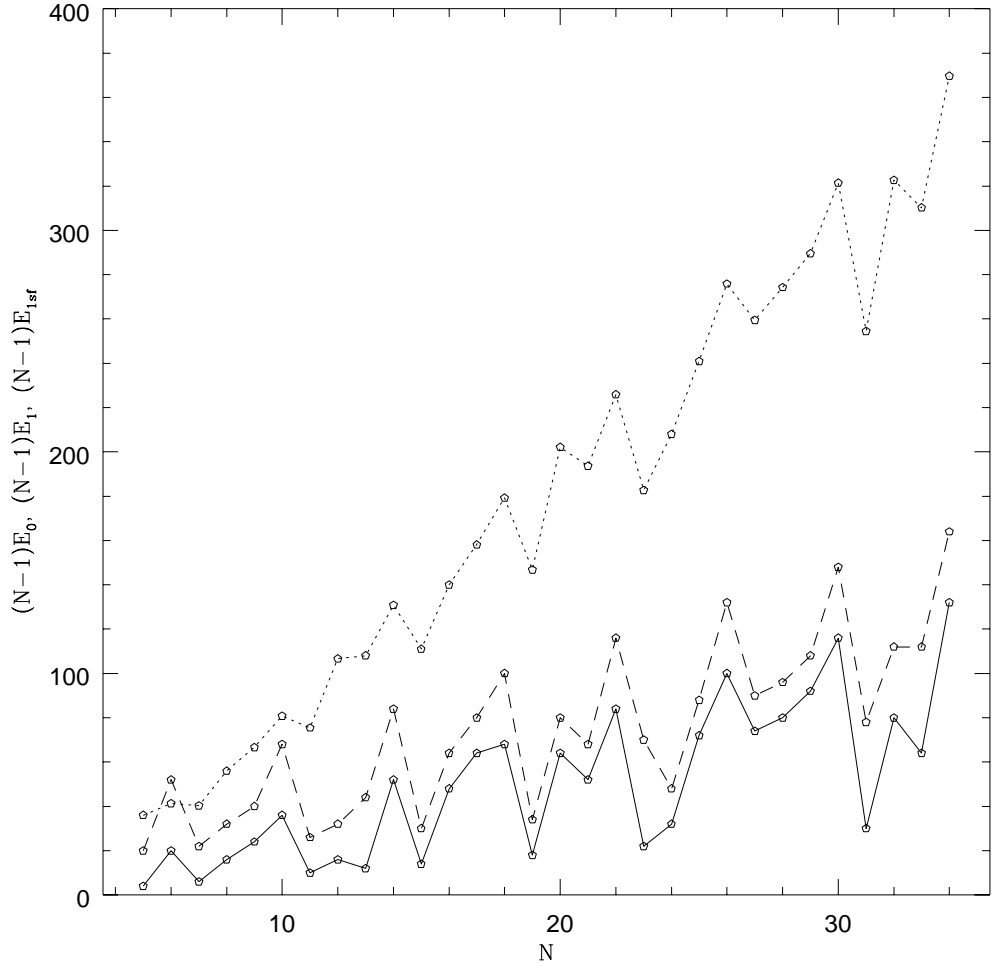}
  \caption[a]{\protect\label{F_SF_ENE}
  $(N-1)$ times the ground state
  energy (continuous curve), first excited state energy (dashes) and
  average energy of configurations obtained by a single spin flip from
  the ground state (dots) as a function of $N$.}
\end{figure}

\item

Few configurations with very small energy start to dominate the
partition function at low $T<T_c$. We note that our estimate for the
constant $A$ coincides with our finite size estimate for the critical
temperature (from the location of the specific heat peak). The
relation $T_c\simeq\frac{1}{A}$ (which holds in the REM
model~\cite{DERRIDA}) seems to apply here with reasonable precision.

This scenario is very similar to the one we are used to see in spin
glasses, when a replica symmetry broken phase exists. In particular it
reminds of Derrida's Random Energy Model (REM) \cite{DERRIDA}, where
at low temperature only a very small set of configurations dominates
the partition function~\cite{DERRIDA,GROMEZ}.

\item

As we can already see from fig. (\ref{F_CVT}) the specific heat
becomes very small in the low temperature region. Very likely it is
exponentially small in the thermodynamical limit. We expect that the
$N^{-1}$ corrections (which in the REM \cite{DERRIDA} are proportional
to $(N(\beta-\beta_c))^{-1}$) dominate the specific heat in the low
temperature phase for $N$ not too large.

\item

Derrida's model does not have the divergence of the specific heat at the
transition point which we have here.  This is likely to be the signature of a
transition of a different nature than the one in Derrida's model.

\end{itemize}

%%%%%%%%%%%%%%%%%%%%%%%%%%%%%%%%%%%%%%%%%%%%%%%%%%%%%%%%%%%%%%%%%%%%%%%%%%
\section{The Golay-Bernasconi Approximation and a First
Replica Computation\protect\label{S_APP}}

Let us try now to give an approximate analytic evaluation of the
thermodynamical properties of the model. We will follow the approach
Golay \cite{GOLAY} has originally introduced (see also Bernasconi work
\cite{BERNAS}) for the open model, and apply it to the periodic
model. We will stress the interest and the obvious limitations of such
a simple approximation (which basically amounts to consider the
correlation functions $C_k$ as independent variables).

Let us consider the periodic model, and the correlation function $C_k$
as defined from eq. (\ref{E_CPERIODIC}). The basic observation is that
on a generic random configuration of the $\s$ the correlation
functions turn out to be also independent variables, randomly
distributed according to a Gaussian distribution with variance
$N$. Therefore for the probability distribution of the correlation
function $C_k$ we can write

\be
  P(C_k) = (2\pi N)^{-\frac{1}{2}}\  e^{-\frac{C_k^2}{2N}}\ ,
  \protect\label{E_GAUSS}
\ee

\noindent
which holds under our statistical independence hypothesis. Here $k$
can vary from $1$ to $N$. Let us take $N$ odd. Since in this case the
correlation functions satisfy the relation

\be
  \protect\label{E_SIMM}
  C_k = C_{N-k}
\ee

\noindent
for all $k$ values, the Hamiltonian (\ref{E_H1}) can be rewritten as

\be
  \protect\label{E_H2}
  H \equiv {2 \over N-1} \sum_{k=1}^{\frac{N-1}{2}} C^2_k\ .
\ee

\noindent
In this case we only need to consider $\frac{N-1}{2}$ modes.  For even
$N$ we should add to $H$ the contribution at $k=\frac{N}{2}$ without
the factor $2$.

In this approximation the partition function is given by

\be
  Z(\beta) = 2^N \prod_{k=1}^{\frac{N-1}{2}}
  \ \{\  \int dC_k\  P(C_k)\  e^{-\beta H(C_k)}\  \}  \ ,
\ee

\noindent
where we have used (\ref{E_SIMM}) and (\ref{E_H2}) to have $k$ running
only up to $\frac{N-1}{2}$. Substituting we get

\be
  Z(\beta) = 2^N \prod_{k=1}^{\frac{N-1}{2}}
  \{ \int \frac{dC_k}{\sqrt{2\pi N}}
  \ e^{-C_k^2(\frac{1}{2N}+\frac{2\beta}{N-1})} \}  \ .
\ee

\noindent For $N$ large we finally get

\be
  \protect\label{E_ZAPP}
  Z(\beta) = e^{ N\  (\ln(2) - \frac{1}{4} \ln(1+4 \beta))}\ .
\ee

\noindent
We have obtained the expression (\ref{E_ZAPP}) under the assumption
that the $\s_i$ are independent variables (and then also the $C_k$
are). This is obviously not true as soon as $\beta > 0$, and
expression (\ref{E_ZAPP}) fails.  Indeed the $C_k$ are not Gaussian
independent random variables. For $\beta > 0$ when evaluating the
partition function we sample the tail of the probability
distribution $P(C_k)$, where the expression (\ref{E_GAUSS}) is not
valid (we will see that the high $T$ expansion does not coincide with
the correct one even at first order). Here we are trying to understand
(until now lacking a better approach: but see later) if at least in a
high-temperature phase eq. (\ref{E_ZAPP}) can be an useful
approximation to the true behavior of our system.

{}From the approximated result for the partition function of
(\ref{E_ZAPP}) we can compute the free energy density (\ref{E_FREE}),
and the usual thermodynamical energy density and entropy. We find

\bea
            f(\beta) &=&  \frac{1}{\beta}
                          (\frac{1}{4} \ln(1+4\beta) -\ln(2) )\ ;\\
\protect\label{E_GBDRES}
            e(\beta) &=&{1 \over 1+4\beta}\ ; \\
  \nonumber s(\beta) &=&   \ln(2) - \frac{1}{4}\ln(1+4 \beta)
                         + \frac{\beta}{1+4\beta}\ .
\eea

\noindent
The behavior of the energy density is quite reasonable, while the
entropy density $s(\beta)$ becomes negative at low temperature (it
goes to $-\infty$ at $T=0$). The entropy density $s(\beta)$ becomes zero at
$\beta_G=10.3702$, where the energy density has the value $e_G \equiv
e(\beta_G)=.02354$.

A possible approximated approach to the problem (along the direction
hinted by Golay and Bernasconi) would be based on saying that this
solution is close to the correct one in the high $T$ phase, for
$\beta\le\beta_G$. One would then claim that a good approximation is
to state that for $\beta > \beta_G$ general thermodynamical properties
(i.e. the fact that both the specific heat and the entropy are not
allowed to become negative) imply that the energy density has to
remain constant

\be
  e(\beta)= e_G\ ,\ \ \forall\  \beta \ge \beta_G\ .
\ee

We have a scenario which is very reminiscent of the REM
\cite{DERRIDA}.  As we have already noticed an obvious drawback of
this point of view, which is built on a series of arbitrary
assumptions, is that it does not reproduce correctly even the first
non-trivial order of the high $T$ series expansion. It captures
however some of the relevant features of the model (like for example
the presence of an abrupt transition at finite $T$), and it seems
worthwhile to try to understand better its features.

Now we will try to apply the replica method to the problem of
sequences with low autocorrelation (at a first stage to try to recover
the results of the Golay-Bernasconi-Derrida (GBD) approximation we
have just discussed). We know that replica methods have been applied
with good success \cite{LIBRO1,LIBRO2} to the analysis of systems
whose behavior has remarkable similarities to the one of our low
autocorrelation sequences. Yet, until now replica approach has been
dealing with system in which quenched randomness plays a major
role. There is nothing of a priori random in our low autocorrelation
sequences, and the replica method could seem here out of place.

However if it is true that the generic properties of the behavior of
low autocorrelation sequences have something to do (at least for not
too low $T$) with the ones of a system with quenched disorder. Then we
can hope to use the replica techniques\footnote{The following
conclusions and the replica computation presented in the next
paragraphs have been obtained independently for the open model
by Jean Philippe Bouchaud and Marc Mezard~\cite{BOUMEZ}.}.

We will want a random system which mimics the properties of our
original ordered system. We will have to identify such a system on the
basis of some general principle, and we will see that this will be
more or less easy in the different cases.

One possible approach is based on considering a Hamiltonian

\be
  \protect\label{E_HRE1}
   H_{\{J\}}(\{\s\})\ ,
\ee

\noindent
which depends on the quenched control parameters $\{J\}$, which are
randomly distributed. For a particular realization of the sequence
$\{J\}$ such random Hamiltonian coincides with our original
Hamiltonian (in the present case with (\ref{E_H2})).  Let us suppose
that we are able to use the replica approach to compute the average of
the thermodynamic functions for the system described by
(\ref{E_HRE1}). Now we can hope that the result obtained for a generic
realization of the random variables $\{J\}$ is the same we would have
obtained by selecting the exact $\{J\}$ sequence which leads to the
original Hamiltonian (\ref{E_H2})). In this case the replica symmetry
gives the correct result for the deterministic model. This way of
reasoning is potentially very dangerous, and can lead to disaster. The
$3d$ Edwards-Anderson model, once understood (for recent progresses
see \cite{MAPARI0}), will not very probably lead to the same solution
of the ferromagnetic $3d$ Ising Model. The issue here deals with how
{\em generic} is the special $\{J\}$ sequence which gives the original
deterministic Hamiltonian, and cannot be solved a priori. A posteriori
for example one can verify if the deterministic and the random models
have the same high-temperature expansion (of course this may lead to
surprises in the low temperature region).

A second possible approach is based on the introduction of a control
parameter $\epsilon$, and of a Hamiltonian

\be
  \protect\label{E_HRE2}
   H_{\epsilon,\{J\}}(\{\s\})\ ,
\ee

\noindent
which interpolates from the random Hamiltonian at $\eps=0$ to the
deterministic Hamiltonian at $\eps=1$. If the interpolation is smooth
and there are no phase transitions in the interval $0 < \eps < 1$ the
perturbative expansion around the result $\eps=0$ (which one should be
able to obtain) could be used to estimate the results for $\eps=1$.

This is the general framework. We hope that, by using one of these
approaches, replica method will enable us to obtain qualitative and
quantitative predictions about the deterministic problem.

Let us start by trying to reproduce the GBD result (i.e.  the simple
approximation we have just studied) in the framework of replica
theory. Our aim will be to consider a soluble random model such that
the probability distribution of correlation functions is Gaussian, as
in (\ref{E_GAUSS}). In the high-temperature phase the free energy
density of such a model should be given by the GBD approximation.

We will consider the Hamiltonian

\be
  \protect\label{E_HBGD}
  H \equiv {2 \over N-1} \sum_{k=1}^{\frac{N-1}{2}} \tC^2_k\ .
\ee

\noindent
In this new model the $\tC_k$ are not simply correlation functions
anymore, but they are given by

\be
  \protect\label{E_GENE}
  \tC_k \equiv \sum_{m,j}J^k_{m,j}\ \s_m \s_j\ ,
\ee

\noindent
where the $J$ are quenched random variables with an average value of
order $\frac{1}{N}$ and variance $\frac{1}{N}$. The precise form of
the distribution is irrelevant. A possible choice for the distribution
of the $J$ variables is

\bea
  J^k_{i,j} = 0   & \mbox{with  probability} & 1 - \frac{1}{N}\ , \\
  \protect\label{E_JUN}
  J^k_{i,j} = 1   & \mbox{with  probability} & \frac{1}{N}\ .
\eea

Random $J$ variables allow connections of random site couples
$i-j$. Since the $\s_j$ are connected randomly it is reasonable to
expect that in the large $N$ limit the modified correlation functions
$\tC_k$ are indeed distributed as independent Gaussian variable. So we
expect our random model defined by (\ref{E_HBGD}), (\ref{E_GENE}) to
have the same behavior (at least in the high-temperature phase) than
the deterministic model defined by (\ref{E_GAUSS}), (\ref{E_H2}).

The model can be studied by means of the usual replica techniques. The
partition function

\be
  Z_{\{J\}}(\beta) = \sum_{\{\sigma\}}
  \ \exp\{ -\frac{2\beta}{N-1}\sum_{k=1}^{\frac{N-1}{2}}\tC^2_k \}
\ee

\noindent is quartic in the spin variables $\sigma$. We can introduce the
variables $X_k$ to disentangle the interaction, getting

\be
   Z_{\{J\}}(\beta) = \sum_{\{\sigma\}} \prod_{k=1}^{\frac{N}{2}}
  \int \frac{dX_k}{\sqrt{8\pi \beta}}
  \exp\{-\frac{X^2_k}{8\beta}
  + \frac{i}{\sqrt{N}} X_k \sum_{m,j}J^k_{m,j}\s_m \s_j ) \}
\ee

\noindent (where for large $N$ we have written $N$ instead of $N-1$).

We want to compute averages over the $\{J\}$ of the free energy density
of the system,

\be
  f(\beta) \equiv \lim_{N\to\infty} (-\frac{1}{\beta N}
  \overline{\ln( Z_{\{J\}}(\beta)) } )\ ,
\ee

\noindent
where the average is a {\em quenched} average over the disorder. We
can employ now the replica trick, rewriting the average over the
disorder of the $\ln Z$ as

\be
  f(\beta) \equiv \lim_{N\to\infty} (-\frac{1}{\beta N}
  \lim_{n\to 0}
  \frac{\overline{Z_{\{J\}}(\beta)^n }-1}{n} )\ .
\ee

\noindent
By adopting the usual abuse of inverting the two limits we finally get

\be
  f(\beta) \equiv \lim_{n\to 0}
  \phi^{(n)}(\beta) \ ,
\ee

\noindent where

\be
  \phi^{(n)}(\beta)
  \equiv \lim_{N\to\infty} (-\frac{1}{\beta N}
  \frac{\overline{Z_{\{J\}}(\beta)^n }-1}{n} )\ .
\ee

Computing the average over the disorder of $Z^n$ is easy. By assuming
a Gaussian distribution for the $J$ variables\footnote{The result of
the computation depends only on the variance of $J$. Imposing a priori
$\langle J \rangle=0$ does not change the result.}
(with zero expectation value and width $\frac{1}{N}$) we find that

\be
  \protect\label{E_ZN1}
  \overline{Z_{\{J\}}(\beta)^n }=
  \sum_{\{\sigma\}} \prod_{k=1}^{\frac{N}{2}}
  \int \prod_{a=1}^{n} ( \frac{dX^a_k}{\sqrt{8\pi \beta}} )
  \exp\{
  - \frac{1}{8\beta}\sum_{a=1}^{n}(X^a_k)^2
  - \frac{1}{2} \sum_{m,j} ( \sum_{a=1}^{n} X^a_k \s^a_m \s^a_j )^2 \} \ .
\ee

\noindent
The second term in the exponential couples the different replicas. We
can rewrite it as

\be
  \frac{1}{2}\sum_{a,b}X_k^a\
  (\sum_m \s_m^a \s_m^b)\ (\sum_j \s_j^a \s_j^b)\  X_k^b\ .
\ee

\noindent
In order to decouple this interaction we write $1$ as

\be
  1 = \prod_{a,b} \int dQ_{a,b}
  \ \delta(\sum_{j=1}^N \s_j^a \s_j^b -NQ_{a,b})\ ,
\ee

\noindent
and use the Lagrange multipliers $\La_{a,b}$ to rewrite the $\delta$
functions

\be
  \protect\label{E_UNO}
  1 = \prod_{a,b}\  [ \int dQ_{a,b}\  d\La_{a,b}\
  \exp\{i\La_{a,b} (\sum_{j=1}^N\s_j^a\s_j^b-NQ_{a,b}) \} ]\ .
\ee

\noindent
Now using eq. (\ref{E_UNO}) in (\ref{E_ZN1}) we can integrate over the
$X_k^a$ variables, and disintegrate the sum over the $\s_j$
configurations. We get

\be
  \protect\label{E_ZN2}
  \overline{Z_{\{J\}}(\beta)^n }=
  \int \prod_{a,b}(dQ_{a,b}) \prod_{a,b}(d\La_{a,b})
  e^{ - N A(\La,Q)  }\ ,
\ee

\noindent where

\be
  \protect\label{E_DEFA1}
  A(\La,Q) \equiv G(Q) + F(\La) + T(\La,Q)\ ,
\ee

\noindent and we have defined

\bea
  G(Q)     & \equiv & \frac{1}{4} \mbox{Tr}
    \ln(\delta_{a,b}+4 \beta Q_{a,b}^2)\\
  \protect\label{E_DEFA2}
  F(\La)  & \equiv & -\ln(\sum_{\{\s\}}\exp\{\sum_{a,b}\La_{a,b}\s^a\s^b \})\\
  \nonumber T(\La,Q) & \equiv & \mbox{Tr} \{ \La_{a,b}  Q_{b,a} \}\ ,
\eea

\noindent
where the trace Tr is taken over the replica indices, and the integral
over $\La_{a,b}$ is taken over the imaginary axis.

In the large $N$ limit $\overline{Z_{\{J\}}(\beta)^n }$ is dominated by its
saddle point value, i.e. we get that

\be
  \phi^{(n)}(\beta) = \frac{1}{\beta} \frac{A_{SP}}{n}\ ,
\ee

\noindent
where by $A_{SP}$ we have indicated the saddle point value of the
expression (\ref{E_DEFA1}).

In the high temperature phase we can look at the replica symmetric
solution, where $Q_{a,b}=0$ for $a\ne b$. The saddle point equations
for $\Lambda$ imply that $Q_{a,a}=1$ (this result is valid at all
temperatures). The expression for the free energy reduces in this way
to eq. (\ref{E_GBDRES}). The result is, as we promised before, the
same of the GBD approximation.

Before studying the properties of the broken replica solution of this
stationary equation, we can get some further insight into the model by
considering the following generalization:

\be
  \protect\label{E_HALPHA}
  H_{(\al)} = \frac{2}{\sqrt{\al} (N-1)} \sum_{k=1}^{\al  N/2} \tC^2_k\ ,
\ee

\noindent
where the quantities $\tC_k$ are defined as in (\ref{E_GENE}).
Here we have only
changed the number of $J$ values which can couple two sites $i$ and
$j$. Since here we are not dealing with pure correlation functions,
but with terms which are coupled or not according to the value of a
random variable, there are no reasons for fixing the total number of
non-zero $J$ values to be of order of $N^2$. For $\al=1$ we recover
our previous model.

The model can be solved for generic $\al$ and one finds results that
are very similar to the previous case. The only difference is that now

\be
  G(Q) \equiv \frac{\al}{4}\mbox{Tr}\ln(1+\frac{4\beta Q^2}{\sqrt{\al}})\ .
\ee

In the limit in which $\al$ goes to infinity all sites are coupled and
the model describes an infinite range $4$-spin interaction.
In this limit one gets

\be
  G(Q) \to \frac{\al^{\frac{1}{2}}}{4}-\frac{1}{8}\sum_{a,b}Q^4_{a,b}\ ,
\ee

\noindent
which is the known result for the $p=4$ model~\cite{GROMEZ}.
For $\al$ going to zero frustration disappears. In other
words the models based on $H_\alpha$ are related to the generic
$4$-spin random models in the same way as the Hopfield models are
related to the Sherrington-Kirkpatrick model.

In the limit $p\to \infty$ the model with a $p$-spin interaction
coincides with the REM \cite{DERRIDA}. In the low temperature phase
replica symmetry is broken at one step \cite{DERRIDA,GROMEZ}. In this
case (where $p\to\infty$) the entropy at the transition and below the
transition point is zero, and the self-overlap parameter $q(1)$ jumps
from $0$ to $1$ at the transition point \cite{GROMEZ}.  Let us also
note that in some sense \cite{GROMEZ} the $p=2$
Sherrington-Kirkpatrick case is a special case, and that as soon as
$p>2$ things change. For example as soon as $p>2$ the phase transition
becomes, as far as the function $q(x)$ is concerned, first order.

We have computed the one step replica broken solution for our
$\al$-dependent model. In this case the matrices $Q$ and $\La$ are
described by the breakpoint $m$ and by their value inside a block. In
the limit $n\to 0$ we find that

\bea
  \protect\label{E_ONESTEP}
  G(Q)       &=& \frac{(m-1)\al}{4m}
    \ln(1+\frac{4\beta}{\sqrt{\alpha}}(1-q^2))
    + \frac{\al}{4m} \ln(1+\frac{4\beta}{\sqrt{\alpha}}(q^2m+1-q^2))\\
  F(\La)     &=& \lambda
    -\frac{1}{m} \ln \int_{-\infty}^{+\infty}\frac{dx}{\sqrt{2\pi}}
      e^{-\frac{x^2}{2}}\cosh^m(\sqrt{2\lambda}x) -\ln(2)  \\
  \nonumber T(\La,Q)   &=&    \lambda q (m-1)  \ .
\eea

We can solve now the saddle point equations for $A_{SP}$ under the
form (\ref{E_ONESTEP}) for the $p=4$ spin interaction (our model for
$\al=\infty$). This gives the free energy density of the one step
replica broken solution (that is exact for the $p\to\infty$
model). Here we find that the entropy at the transition is very small
(about $.01$) and that the self-overlap parameter $q(1)$ is very close
to $1$ (it is greater than $.95$).  The GBD approximation describes a
scenario with a zero entropy at the transition point, and $q(1)$
jumping from $0$ to $1$. That means that the difference between the
GBD approximation and the infinite range $4$ spin interaction is of
the order of a few percent (on the expectation values of typical
thermodynamical observables).

The situation improves if we look at to our model with $\al=1$. In
this case, assuming one step replica symmetry breaking, we find that
the entropy at the transition is tiny (smaller than $.0001$) and that
the self-overlap parameter $q(1)$ is very close to $1$ (it is greater
than $.99$). The inverse transition temperature is practically
identical to the one we have found in the GBD approximation (after
eqs. (\ref{E_GBDRES})).  Here the Golay-Bernasconi-Derrida approximation
is practically perfect.

That completes a quite detailed understanding of our
$\alpha$-dependent disordered model. We have obtained the one step
replica broken solution of the model, and it has been useful to show
that the model undergoes a finite $T$ phase transition to a glassy
region, where the partition function is dominated by a restricted set
of states. The corrections to the GBD approximation can be computed
and they turn out to be very small.

%%%%%%%%%%%%%%%%%%%%%%%%%%%%%%%%%%%%%%%%%%%%%%%%%%%%%%%%%%%%%%%%%%%%%%%%%%
\section{The High-Temperature Expansion of the Low
Autocorrelation Model and a Hartree-Fock Resummation\protect\label{S_HIG}}

In the previous section we have used replica theory to analyze and
solve a model which does not have the same high-temperature expansion
than the low autocorrelation model we started from, i.e. the one
defined from (\ref{E_H1},\ref{E_CPERIODIC}). Altogether we have been
acting quite recklessly.  We have introduced a (maybe not so good)
approximation to our original deterministic model, and we have defined
(in (\ref{E_HALPHA},\ref{E_GENE})) and solved a model with quenched
random disorder which reproduces such an approximation. This has been
useful to show that replica theory can play an important role even in
the understanding of statistical models which do not contain quenched
disorder in their formulation. Still, now we are interested in
stepping forward, and getting a deeper understanding of our original
model.

The first tool we will use to learn more about the full low
autocorrelation sequence model is the high-temperature expansion. As
matter of principle this can be done in a very straightforward way,
but on practical grounds the fact that model is non-local creates lot of
complications. For example the coefficient of the high $T$ expansion
(of the energy density, let us say) are not polynomial in $N$, as they
would be for a well behaved interaction. Only the leading contribution
(in $N^{-1}$) at each order in $\beta$ is universal, while subleading
corrections tend to depend on the cardinality of $N$ (for example we
can have a given polynomial for odd $N$ and a different one for even
$N$, and so on with more and more complicate behaviors).

The direct evaluation of the high-temperature approximation in
$x$-space is possible, but not very convenient, because of the
problems we have just described. We have just used it to check the
general behavior of particular classes of diagrams. We have found
convenient to use instead the momentum space representation
Hamiltonian (\ref{E_HF1}). We have computed the leading terms in
$N^{-1}$ of the first $3$ non-trivial $\beta^{-k}$ expansion
coefficients for the free energy density, i.e we have only considered
connected diagrams in the expansion of the partition function
$Z(\beta)$.

For example the coefficient of the $\beta^2$ term (for the free energy
density) is

\be
  \protect\label{E_COE3}
  \lim_{N\to\infty} \frac{1}{3!} \frac{1}{N^7}\  [\sum_{k_1,k_2,k_3}
  |B(k_1)|^4|B(k_2)|^4|B(k_3)|^4 ]_c\ ,
\ee

\noindent
where the small $c$ signifies that we have only included in the sum
contributions from connected diagrams. In order to compute the
diagrams\footnote{One has to be be careful in noticing that
$|B(p)|=|B(-p)|$ in order to avoid double counting.} one has to
analyze separately the case where $k_1 = k_2 = k_3$, the case where
two $k_i$ are equal and the one where all the three $k$'s are
different. By using this approach we have been able to find that the
first $3$ orders of the small $\beta$ expansion of the energy density
(deduced from the free energy density by the usual relation $ e(\beta)
= -\partial(\beta f(\beta))/\partial\beta$) are given by

\be
  e(\beta) = 1 - 8 \beta + 160 \beta^2 + O(\beta^3)\ .
\ee

We have also looked at subleading contributions to the energy density
$\beta^2$ term, both in real space and in momentum space. One easily
sees that there are in this case diagrams which are proportional to

\be
  \frac{1}{N^3} \sum_{k_1} \de_N (3k_1)\ ,
\ee

\noindent
where $\de_N(k)=1 \iff k=0 (\mbox{mod}\ N)$. A term of this kind gives
a non-zero contribution only if $N$ is multiple of $3$.

The number of relevant diagrams proliferates at the next order in
$\beta$ ($O(\beta^{3}$ for the internal energy). Here subleading
corrections also contain terms proportional to

\be
  \frac{1}{N^4} \sum_{k_1} \de_N (5k_1)\ ,
\ee

\noindent
which now also distinguish the $N$ values which are multiple of $5$.

At last we have been able to check that at order $\beta^4$ (again for
the internal energy) there are terms of order $N^{-5}$ which even for
$N$ odd have a different expression depending on if $N=1 (\mbox{mod}
4)$ or not.

Indeed the easiest way to compute the high-temperature expansion
coefficients turned out to be based on the exact solution of the
systems with size up to $N=38$ we have described before (together with
the insight about the diagram structure we have described in the
former paragraphs). We have used here the density of states
$\cN_N(E)$. The cumulant of order $k$

\be
  \langle H^k\rangle_c^{(N)}(\beta=0)
\ee

\noindent
can be used indeed to fit the $N^{-j}$ coefficients of the $\beta^k$
term in the high-temperature expansion. In better educated models such
coefficients would be simple polynomial in $N$, and the information we
have would (for $N$ up to $38$) would allow us to fit a large number
of terms. Here on the contrary we have a polynomial behavior only on
selected subsequences of $N$ values (that we have discussed
before). So the number of terms we have been able to work out is quite
low.

Already the term of order $1$ in the energy density is different for
odd and even $N$ values. We find that

\bea
  e_0^{(o)}(\beta,N) &  = & 1-\frac{1}{N}\ ,\\
  \protect\label{E_E0BN}
  e_0^{(e)}(\beta,N) &  = & 1 \ ,
\eea

\noindent
where by the subscript to $e$ we indicate the order in $\beta$, and by
the upperscripts $(e)$ and $(o)$ we indicate respectively even and
odd.  The same structure survives at next order in $\beta$, giving

\bea
  e_1^{(o)}(\beta,N) &  = & \beta ( - 8 + \frac{24}{N} - \frac{16}{N^2})\ ,\\
  \protect\label{E_E1BN}
  e_1^{(e)}(\beta,N) &  = & \beta ( - 8 + \frac{32}{N^2})\ .
\eea

\noindent
We have been able to check directly from the diagrammatic expansion
the full expressions (\ref{E_E0BN}) and (\ref{E_E1BN}) (including all
subleading correction).

We have already explained that at order $\beta^2$ we get different
results depending on if $N$ is multiple of $3$ or not. For $N$ of the
form $3n+1$ and $3n+2$ (integer $n$) we find

\be
  \protect\label{E_E2BNA}
  e_2^{(\overline{3})}(\beta,N)   =  \beta^2 ( 160
  - \frac{1008}{N} + \frac{1856}{N^2} - \frac{1008}{N^3})\ ,
\ee

\noindent
while for $N$ multiple of $3$ we get

\be
  \protect\label{E_E2BNB}
  e_2^{(3)}(\beta,N)   =  \beta^2 ( 160
  - \frac{1008}{N} + \frac{1856}{N^2} - \frac{752}{N^3})\ ,
\ee

\noindent
where here by the superscripts $(3)$ and $(\overline{3})$ we have
designated $N$ values which are and are not multiple of $3$.

At next order in $\beta$ ($\beta^3$ for the internal energy density)
we have been able to find the exact polynomial only for $N$ non
multiple of $3$ and $5$ (that for our $N$ values, and indeed up to
$N=77$, coincide with prime values).  Here we had $9$ numbers (the
momenta for primes going from $7$ to $37$) and $5$ coefficients to
find. That is redundant enough to allow to check carefully that we did
the right thing. For the other $N$ value subsequences at this order,
and next orders in $\beta$, we have not been able to calculate the
expansion coefficients. Here we find (with obvious notation)
\be
  \protect\label{E_E3BN}
  e_3^{(\overline{3},\overline{5})}(\beta,N)   =  \beta^3
  (1-\frac{1}{N}) ( -5248
  + \frac{43520}{N} - \frac{124672}{3N^2} - \frac{781312}{3 N^3})\ .
\ee

As far as the leading $N^{-k}$ term is concerned we have in this way
gained one order in our small $\beta$ expansion, by finding

\be
  \protect\label{E_HIGH_ALL}
  e(\beta) = 1 - 8 \beta + 160 \beta^2 - 5248 \beta^3\ + O(\beta^4).
\ee

Can we learn something more about the model in its high-temperature
phase? We hope so, and in order to do that we will now try to write
down a statistical model that hopefully resums the high-temperature
expansion.

But for a trivial shift in the energy we can rewrite the partition
function of the low autocorrelation model as

\be
  Z(\beta) = \int \prod_{i=1}^N [d\s_i]
  \exp\{-\frac{2\beta}{N}\sum_{p=1}^{\frac{N}{2}}|B_p|^4   \} \ ,
\ee

\noindent
where the $B_p$ are the Fourier transformed variables defined in
(\ref{E_FT}). Let us define now the new Hamiltonian

\be
  \protect\label{E_HNU}
  H_\nu(B) \equiv  \frac{2}{N} \sum_{p=1}^{\frac{N}{2}}|B_p|^{2\nu} \ ,
\ee

\noindent
where now the $B_p$ are the fundamental variables of the model.
In the case $\nu=2$ this Hamiltonian coincides (apart from the trivial
energy shift) with the one of the original model. The case $\nu=1$
will be of large importance, since in this case $H_1=N$ for all
$\{\s\}$ configurations.

We can obtain a very simple result if we select only the
contributions to the high $T$ expansion which come from diagrams in
which all momenta are set to be equal. That means for example we choose
from (\ref{E_COE3}) only contributions with $k_1=k_2=k_3$.

It is easy to resum these diagrams.  In this case we find that the
probability distribution for the $B_p$ factorizes in an independent
contribution for each momentum, and we get that

\bea
  \langle f(B)\rangle & \equiv & \int \dbbb\   f(B) P(B)\ ,\\
  P(B) & \propto & \exp( -|B|^2 - \beta |B|^\nu)\ .
\eea

\noindent
This result cannot be the correct, complete answer, since it implies
that $|B|^2$ is a function of $\beta$, while we know that for all
$\beta$ values the correct answer is

\be
  \langle |B|^2 \rangle= \langle\sigma ^2 \rangle =1\ .
\ee

\noindent
However we will see with pleasure that we are not very far from the
correct answer.

It is clear that leading contributions coming from diagrams where the
flowing momenta are different exist, and we will have to consider
them. These contributions generate an interaction in our effective
Hamiltonian, and they cannot be neglected. A detailed inspection of
the large $N$ leading contributions in the high temperature expansion
leads us to conjecture that for large $N$ the partition function of the
low autocorrelation model can be written (at least in the high $T$
phase) as

\be
  \protect\label{E_HARFOC}
  Z_\nu(\beta) \equiv
  \int \prod_{p=1}^{\frac{N}{2}} [ dB_p \ d\overline{B}_p ]
  \exp \{ -\sum_{p=1}^{\frac{N}{2}} |B_p|^2 \}
  \exp \{\frac{N}{2} g(\cD) \}
  \exp \{ -\beta \sum_{p=1}^{\frac{N}{2}} H_\nu(B)  \}\ ,
\ee

\noindent
where the operator $\cD$ is defined as

\be
  \cD \equiv \frac{2}{N} \sum_{p=1}^{\frac{N}{2}}
  \frac{\partial^2}{\partial B_p \partial \overline{B}_p}\ ,
\ee

\noindent
the integral is taken over real and imaginary part of $B_p$, and $g$
is a function which {\em does not} depend on $\nu$ and which we will
explicitly compute.

We have here a guess for the form of $P(B)$. We have a Gaussian weight
over the $B$'s, a weight given by the Hamiltonian and an interaction
correction term, the function $g$. Such a conjecture comes from a
comparison with the dominant contributions in the high-temperature
expansion of the original formulation of the model. For all terms we
have been able to think about the correspondence holds\footnote{The
doubtful reader will find a different derivation of this result in
section (\ref{S_REP}). }. As we shall see later the expression we have
conjectured essentially corresponds to a Hartree-Fock approximation.

Let us start by evaluating the partition function (\ref{E_HARFOC}) for
a generic function $g$.

As usual it is convenient to introduce the representation

\be
  1 = \int dx \ \delta (x-{\cD}) =
  \int dx \int d\la\  \exp \{ i\la(x -{\cD}) \} \ .
\ee

\noindent
By inserting the $\delta$ function the $\dbbb$ integrals factorize, and we get

\be
  Z_\nu(\beta) = \int dx \int_{i{\bf R}} d\lambda\
  e^{\frac{N}{2}(\lambda x + g(x))}
  [\dbbb e^{-|B|^2}
  e^{-\lambda \frac{\partial^2}{\partial B \partial \overline{B}}}
  e^{-\beta|B|^{2\nu}}]^{\frac{N}{2}}\ ,
\ee

\noindent
where the derivative operator only acts on the last exponential function.

In order to compute $Z_\nu(\beta)$ we can use now the familiar expression for
the heat kernel. Let us consider the real variable $z$, and the operator $O$
acting on functions $f$. The kernel of $O$, $K_O$, is defined as

\be
  (Of)(z) = \int dz'\ K_O(z,z') f(z')\ .
\ee

If we consider now the operator
$\exp\{-\lambda\frac{\partial^2}{\partial z^2}\}$
we find that its kernel (the heat
kernel) has the form

\be
  \frac{1}{2\sqrt{\pi \lambda}} e^{-\frac{(z-z')^2}{4\lambda}}\ .
\ee

\noindent
We can use this last formula to rewrite $Z_\nu(\beta)$ (the most transparent
approach consists in using as independent variables real and imaginary part of
$B$, getting in this way two real heat kernels). Now the integrals over the
left variable of the two kernels are Gaussian. After integrating them out we
are left with the expression

\be
  Z_\nu(\beta) = \int dx \int_{i{\bf R}} d\lambda \
  e^{\frac{N}{2}[\frac{x}{4\mu} + \tilde{g}(x)+\ln(\mu)+L(\beta,\mu)]}\ ,
\ee

\noindent
where we have defined $\mu\equiv (1+4\lambda)^{-1}$, $\tilde{g}(x) \equiv g(x)
- \frac{x}{4}$, and

\be
  L(\beta,\mu) \equiv \ln\{\int \dbbb e^{-\mu|B|^2-\beta|B|^{2\nu}}\}\ .
\ee

The former expression can be evaluated in the large $N$ limit by taking its
saddle point. One finds that

\bea
  \tilde{g}'(x) + \frac{1}{4\mu} &=& 0 \ ,\\
   -\frac{x}{4\mu^2} +  \frac{1}{\mu} - \langle |B|^2 \rangle_{eff} &=& 0\ ,
\eea

\noindent
where the expectation value is computed with the effective local
Hamiltonian:

\be
{\cH}(B) \equiv \mu |B|^2 + \beta |B|^{2\nu}\ .\protect\label{E_LOCAL}
\ee

We have also to impose that the sum of the $|B|^2$ is one, which was a
crucial feature of our original model. If the expectation value of
$|B|^2$ is one than the expectation value over the effective
Hamiltonian also has to be one. That gives us a third equation

\be
  \protect\label{E_B2UNO}
  \langle |B|^2 \rangle_{eff} = 1\ .
\ee

We have found that the saddle point free energy is determined from

\bea
  \tilde{g}'(x) + \frac{1}{4\mu} &=& 0 \ ,\\
   -\frac{x}{4\mu^2} +  \frac{1}{\mu}  &=& 1\ ,\protect\label{E_SADDLE} \\
   \langle |B|^2 \rangle_{eff} &=& 1\ . \nonumber
\eea

\noindent
The second of equations (\ref{E_SADDLE}) gives us $x$ as a function of
$\mu$, i.e.

\be
  x = 4(\mu-\mu^2)\ .
\ee

\noindent
Now we can use the first of equations (\ref{E_SADDLE}) to determine
the function $\tilde{g}$. We find that

\be
  \tilde{g}'(x) = - \frac{1}{2(1+(1-x)^{\frac{1}{2}})}\ ,
\ee

\noindent
that gives

\be
  \tilde{g}(x) = - \ln(1+\sqrt{1-x}) + \sqrt{1-x}
\ee

\noindent
(where we have omitted an irrelevant constant).

Now it easy to compute the saddle point free energy density. One only
has to use the third of equations (\ref{E_SADDLE}) to determine the
saddle point value of $\mu$. The expression for $\ln(Z_\nu(\beta))$
eventually greatly simplifies.

If we are only interested in computing the expectation value of the
energy density we can use a shortcut, by noticing that the energy
density of the model is the derivative with respect to $\beta$ of the
logarithm of the partition function, and can be expressed as

\be
  e(\beta) = \langle |B|^{2\nu}\rangle_{eff}\ .
\ee

\noindent
The former identity has to be supplemented by the condition
(\ref{E_B2UNO}), i.e. $\mu$ is fixed by setting the expectation value
of $|B|^2$ over the effective Hamiltonian to one.  In a language
suitable to field theory addicts we can say that only tadpole diagrams
have survived.  The total contribution of the tadpoles is fixed by the
condition eq. (\ref{E_B2UNO}).  Given the simplicity of the result it
is quite likely that our proof may be simplified.

We have tested the correctness of our conjecture by computing the
corresponding high-temperature expansion and by verifying that the
first $4$ coefficients are indeed correct, and coincide with
(\ref{E_HIGH_ALL}). Our Hartree-Fock resummation is equivalent, as far
as we can see, to the complete low autocorrelation model at least in
the whole high $T$ phase.

%%%%%%%%%%%%%%%%%%%%%%%%%%%%%%%%%%%%%%%%%%%%%%%%%%%%%%%%%%%%%%%%%%%%%%%%%%
\section{The Replica Approach\protect\label{S_REP}}

In the previous section we have succeeded to write a closed form for
the solution of our model in the high $T$ phase. We are ready now to
try to achieve the main result of this paper, and show that replica
theory can be used to obtain the solution of a non-random spin
model. We will define a disordered model which has the correct
high-temperature expansion of the initial non-random model (and
contrary to the GBD case we will not need here any approximation), and
that can be solved at all temperatures by using the replica method.

The model we propose is based on the simple observation that the Fourier
transform is a very special unitary operator.  Naively one could think to write
a model where the Hamiltonian is the one defined in (\ref{E_HNU}) with $\nu=2$,
but the basic configurational variables which will be integrated over are

\be
  B(p) \equiv \sum_j U_{p,j}\s_j\ ,
\ee

\noindent
where the $U$ matrix are generic unitary transformations, and compute the
thermodynamic properties of the model for a random choice of the $U$ matrices.
The point is here that the Fourier transform is {\em one particular} unitary
transformation, and we try to understand what happens if we substitute it with
{a random} transformation.

One has to be slightly more sophisticated than that, since by using generic
unitary matrices $U$ already at the first orders of the high $T$ expansion
one gets a result that is different from the one one obtains when using the
Fourier transform. This effect can be traced to the fact that by using a
generic unitary transformation we are ignoring the fact that in the original
model we were transforming real functions, and there

\be
  \protect\label{E_SIMM_P}
  B(p)= \overline {B(-p)}\ .
\ee

\noindent
This reality property turns out to be crucial, and our model with
quenched disorder will have to account for it. In order to satisfy
this constraint we will consider the Fourier transform as an
orthogonal transformation which carries a real function in a complex
one, which satisfies (\ref{E_SIMM_P}).  We introduce the variables
$A(p)$ by

\bea
  B(0)           &=& A(1)\ , \\
  B(\frac{N}{2}) &=& A(N)\ , \nonumber \\
  B(p) &=& A(2p) + i A(2p+1) \ \mbox{\rm for}\  p=1,\frac{N-1}{2}\ ,
\eea

and (for even $N$) rewrite the Hamiltonian (\ref{E_HNU}) as

\be
  H =\sum_{p=1}^{\frac{N-1}{2}} |A(2p)+iA(2p+1)|^4 +|A(1)|^4+|A(N)|^4
\ee

\noindent
Our random model will be defined, in the large $N$ limit, from the
equivalent Hamiltonian (we are forgetting contributions of relative
order of magnitude $N^{-1}$)

\be
  \protect\label{E_HRA1}
  H \equiv \sum_{p=1}^{\frac{N}{2}} |A(2p-1)+i A(2p)|^4\ ,
\ee

\noindent
where the $A$ variables are defined from the spin variables $s_j$ as

\be
  \protect\label{E_VAR1}
  A(p) \equiv \sum_{j=1}^{N} O_{p,j}\s_j\ ,
\ee

\noindent
and the $O_{p,j}$ are random orthogonal transformations, over which we will
integrate.

The model we have obtained can be studied using the replica
approach. In order to present the replica computation for models of
this kind in a compact way we will describe the solution of a model
based on unitary matrices. An explicit computation shows that if we
solve the orthogonal model (\ref{E_VAR1}) along the same lines we
obtain (apart from a rescaling of $\beta$) the same thermodynamical
behavior in the large $N$ limit.  We define the Hamiltonian

\be
  \protect\label{E_HRA2}
  H \equiv \sum_{p=1}^{\frac{N}{2}}|C(p)|^{2\nu} \ ,
\ee

\noindent
where

\be
  \protect\label{E_VAR2}
  C(p) \equiv \sum_{j=1}^{\frac{N}{2}} U_{p,j}\ \tau_j\ ,
\ee

\noindent
the $U$'s are random unitary transformations and $\tau_j\equiv
\s_{2j-1}+ i \s_{2j}$. We have effectively written a model which is
based on $\frac{N}{2}\times\frac{N}{2}$ unitary matrices (naively we
would have used $N \times N$ unitary matrices), ensuring in this way
to get the correct normalization of the free energy in the high $T$
expansion.  The aim of this section will be to solve this model (which
will eventually be of interest for us for $\nu=2$) and to show that
its high-temperature expansion is the same than for the original low
autocorrelation sequence model.

We proceed as in section (\ref{S_APP}) and introduce replicas. We find that

\bea
  \overline{Z^n} &\propto&  \int dU d\la dC \overline {d\la dC} \\
                 &\sum_{\{\tau\}}&
  \exp \{ \sum_a [ \sum_{p=1}^{\frac{N}{2}} |C^a(p)|^{2\nu}
  +(i(\la_a(p)C^a(p) -\la_a(p) \sum_{j=1}^{\frac{N}{2}} U_{p,j}\tau_j^a)
  + h.c.)]\}\ ,
\eea

\noindent
where with $d\la$ and $dC$ we indicate respectively
$\prod_{a,p}\la_a(p)$ and $\prod_{a,p}C_a(p)$, with $(a=1,n)$,
$(p=1,\frac{N}{2})$. The integrals are taken over the real and
imaginary parts of the variables $\la$ and $C$, and the integral over
$dU$ is over the unitary group. We have to compute an integral of the
form

\be
  \label{E_INTORT}
  \int dU \exp\{\sum_{p,j=1}^{\frac{N}{2}}\Omega_{p,j} U_{p,j}+h.c.\}\ ,
\ee

\noindent
with $\Omega_{p,j}=\sum_a\,\la_a(p)\tau_j^a$ ,
$\mbox{\rm Tr}(\Omega U)\sim N$, and the
integral is performed over the unitary group.  This problem has been solved
in full generality by Brezin and Gross \cite{BREGRO}. However their formula is
more complicated of what we need here. At finite non-zero $n$, in the limit of
$N$ going to infinity, only the terms containing one single trace operation
survive, and the integral is given by

\be
  \int dU \exp \{ \mbox{Tr} (\Omega U + h.c.) \}
  = \exp \{ \frac{N}{2} \mbox{Tr} G({ \Omega \Omega^*\over N^2}) \}\ ,
\ee

\noindent
where $G(z)$ is a function which form we want to derive. Let us consider the
case in which the matrix $\Omega$ has one  single element different from zero,
for example $\Omega_{11}\equiv \frac{zN}{2}$.
We define the function $G(z)$ from the relation

\be
  I \equiv \int dU \exp \{\frac{N z U_{1,1}}{2}+h.c.\}
  = \exp\{N  G(\frac{|z|^2}{4})\}\ .
\label{E_INTORT2}
\ee

\noindent
The integral over the unitary group is given by

\bea
  I &=&       \int dU \exp (\frac{N}{2} z U_{1,1}+h.c.) \\
    &=&       \int dx\  \delta (\sum_{j=1}^{N}x^2_j -1)
              \exp \{\frac{N}{2} z x_1\}\\
    &\simeq&  \int dx_1 (1-x_1^2)^{\frac{N}{2}}
              \exp \{\frac{N}{2} z x_1\}\ . \nonumber
\eea

\noindent
We have used here the fact that a randomly chosen line of the unitary
matrix is only constrained to have the sum of its elements equal to
one.  The last integral can be evaluated by using the saddle point
method. We find

\bea
I  &\sim& \int dx_1 \exp(\frac{N}{2} f(x_1,z))\\
   &=&    \int dx_1 \exp(\frac{N}{2}(\log(1-x_1^2)+z x_1))\ .
\eea

\noindent
The stationary point $x_0$ of this saddle point equation gives
$I\sim\exp(\frac{N}{2}f(Z))$. Using eq.(\ref{E_INTORT2}) we find

\be
  G(z) = - \ln(\sqrt{1+z}+1) + \sqrt{1+z}\ .
\ee

\noindent
This result can also be derived using the Brezin and Gross formulae
\cite{BREGRO}. $G$ corresponds to the function $\tilde{g}$ of  the
previous section.

Now we have to compute $\mbox{\rm Tr} G(\frac{\Omega \Omega^*}{N^2})$.
It is easy to verify that for all positive integer values of $P$

\be
  \mbox{\rm Tr} \{ (\frac{\Omega \Omega^*}{N^2})^P \} =
  \mbox{\rm Tr} \{ (\La Q)^P \}\ ,
\ee

\noindent
where $\La$ and $Q$ are $n \times \ n$ matrices, defined as

\bea
  \La_{a,b} &=& \frac{1}{N}
    \sum_{p=1}^{\frac{N}{2}} \la(p)^a \overline{\la(p)^b}\ , \\
  Q_{a,b} &=& \frac{1}{N}
    \sum_{k=1}^{\frac{N}{2}} \tau^a_k \overline{\tau^b_k}\ .
\eea

\noindent
That implies

\be
  \mbox{\rm Tr}  G({\Omega \Omega^* \over N^2}) =
  \mbox{\rm Tr}  G(\La Q)\ .
\ee

The computation now continues using the standard techniques introduced
in the section $4$.  First we introduce auxiliary fields $R$ and
$M$ associated to the matrices $Q$ and $\La$ respectively:

\bea
  1 =  \prod_{ab}\int dQ_{ab}
       \delta(Q_{ab}-\frac{1}{N}\sum_i\tau_i^a\overline{\tau_i^b})\\
    =  \prod_{ab}\int dQ_{ab} dR_{ab} \exp\{iR_{ab}(Q_{ab}-
       \frac{1}{N}\sum_i\tau_i^a\overline{\tau_i^b})\} \ ,\\
\eea

\noindent
and analogously for $\La_{ab}$ and the Lagrange multipliers $M_{ab}$

\bea
  1 =     \prod_{ab}\int d\La_{ab}
          \delta(\La_{ab}-\frac{1}{N}\sum_p\la(p)^a\overline{\la(p)^b})\\
    \sim
          \prod_{ab}\int d\La_{ab} d\tilde{M}_{ab}
          \exp\{i\tilde{M}_{ab}(\La_{ab}-
          \frac{1}{N}\sum_p\la(p)^a\overline{\la(p)^b})\}\ .\\
\eea

\noindent
Putting all together we find that we need to compute

\bea
  \overline{Z^n} &\sum_\tau&
  \int d\lambda d\overline{\lambda} dC d\overline{C}
  \int d\La d\tilde{M} dQ dR\\
  &\exp\{&i \tilde{M} (\La-\frac{1}{N} \sum_p \lambda_p \overline{\lambda_p}\}
  \exp\{i \tilde{R} (Q-\frac{1}{N} \sum_i \tau_i \overline{\tau_i}\}\\
  &\exp\{&N G(\La Q)\} \nonumber
  \exp\{|C|^{2\nu}\}
  \exp\{i\lambda C + h.c.\}\ .
\eea

\noindent
Performing the integration over the $\lambda $ variables we finally
obtain that $(N \ln(Z))^{-1}$ is given by the stationary point of

\be
  \overline{Z^n}=\int dQ dR d\La dM\  \exp\{N A[R,Q,\La,M]\}\ ,
\ee

\noindent

(where we have defined $M\equiv (4\tilde{M})^{-1}$) which means

\be
  \beta f = -\frac{1}{Nn}\,A_{SP}[R,Q,\La,M] \ .
\ee

\noindent
The function $A$ is given by

\bea
  &A[R,Q,\La,M]& = \\
    &F_f(M)& + \mbox{\rm Tr} \ln(M)
    - \mbox{\rm Tr} \frac{\Lambda}{4M} +\mbox{\rm Tr} G(\Lambda Q)
    - \mbox{\rm Tr} R Q + F_s(R)\ ,
\eea

\noindent
where

\bea
  \exp \{ F_f(M) \} &=& \int \dccc\  \exp \{ -\beta \sum_a |C_a|^{2\nu}
    -\sum_{a,b}C_aM_{a,b}\overline{C_b} \}\ ,\\
  \exp \{ F_s(R) \} &=& \int d\{\tau\}
    \exp \{\sum_{a,b}R_{a,b}\tau_a\overline{\tau_b}\}\ .
\eea

\noindent
The previous formula is also valid in the case of a continuous distribution
of the spins $\sigma$. In the present case the spin take the discrete values
$\pm 1$, and we have to substitute the integral by a sum.

In order to solve the saddle point equations we start by  eliminating some of
the auxiliary variables. The full set of saddle point equations for $A$ gives:

\bea
\frac{\partial A}{\partial R_{ab}}
  =-Q_{ab}+ \frac{\partial F_s(R)}{\partial R_{ab}}   =0\ ,\\
\frac{\partial A}{\partial \La_{ab}}
  =-(\frac{1}{4M})_{ab}+(QG'(\La Q))_{ab}=0\ ,\\
\frac{\partial A}{\partial Q_{ab}}
  =-R_{ab}+(\La G'(\La Q))_{ab}=0\ ,\\
\frac{\partial A}{\partial M_{ab}}
  =\frac{\partial F_f(M)}{\partial M_{ab}}
+(\frac{1}{M})_{ab}+(\frac{\La}{4M^2})_{ab}=0\ ,
\eea

\noindent
After some algebra and using the relation

\be
  G'^2(z)=\frac{1}{4z}-\frac{G'}{z}
\ee

\noindent
we can phrase our result in a very simple form. The free energy is given by the
stationary point of

\be
  A[M,R]= F_f(M)+ F_s(R) +\mbox{\rm Tr}\ln(4(M-R))\ .
\ee

\noindent
The expectation values of quantities which are local in momentum or in
configuration space can be computed using respectively the simple
Hamiltonians

\bea
  {\cal H_M} &\equiv& -\beta \sum_a
  |C_a|^{2\nu}-\sum_{a,b}C_aM_{a,b}\overline{C_b}\ ,\\
  {\cal H_R}&\equiv&
  \sum_{a,b}R_{a,b}\tau_a\ba{\tau_b} \ .
\eea

\noindent
The saddle point equations for the stationary free energy are now

\bea
  \langle C_a\overline{C_b}\rangle_M=
  \langle \tau_a\overline{\tau_b}\rangle_R=
  Q_{a,b}\\ (M-R)\  Q =1\ ,
\eea

\noindent
where the mean values $\langle...\rangle_M$ and $\langle...\rangle_R$
are evaluated using the Hamiltonians $\cal{H_C}$ and $\cal{H_R}$
respectively.  The first condition is a clear consequence of the
unitarity of the transformation. The second equation has a less clear
meaning\footnote{We feel a bit guilty of presenting such a complicated
proof for such a simple results, but this is the best we have been
able to do.}.

In the high temperature phase the different matrices are non-zero only
in their diagonal part. This can be computed in the annealed case
$n=1$. In this case the different matrices have a unique element
$Q=1$, $\La=\lambda$, $R=r$ and $M$. The free energy is given by:

\be
  \beta f=-\ln\int \dccc\  \exp\{-\beta|C|^{2\nu}-M|C|^2\} - \log(2)\ ,
\ee

\noindent
where $M$ is determined by the simple equation

\be
  \langle |C|^2\rangle_M=1\ .
  \label{eqc}
\ee

\noindent
The internal energy is given by the relation

\be
  u(\beta)=\frac{\partial (\beta f)}{\partial \beta}=\langle
  |C|^{2\nu}\rangle_M\ ,
  \label{equ}
\ee

\noindent
which coincides with the corresponding equations of the previous
section a part from a rescaling of $\beta$.

We have shown that our model reproduces the high-temperature expansion of
the effective action conjectured in the previous section. For a random system
it is well known that the annealed free energy is a lower bound to the quenched
free energy. That enables us to develop at least a partial analysis of our
results without doing the explicit computation of the replica
symmetry breaking in the limit $n\to 0$. Let us notice, indeed, that in this
light the results of the previous section imply that the ground state energy
of the model is greater than $0.025$.

Explicit formulae can be written in the case of one step replica
symmetry breaking. We want all the three matrices $R,Q,M$ to
commute. To this end we break each one of these matrices into
sub-blocks of equal size $m$. The different elements are, for instance
in case of the matrix $M$, $M_{aa}=M_D$ and $M_{ab}=M_1$ if the
indices $(a,b)$ do belong to the same sub-block of size $m$, while
otherwise $M_{ab}=0$. The same holds for the matrix $R$.  The
variational parameters are now $m, M_D, M_1, R_D, R_1$ and $q_1
(Q_{aa}=2)$, and the saddle point equations are:

\bea
  \int d\mu(z) \langle |C|^2 \rangle_z &=& 2\ , \\
  \int d\mu(z) |\langle C \rangle_z|^2=
  \int d\rho(h)|\langle \tau \rangle_h|^2 &=& q_1\ , \\
  (M_D-R_D)+(m-1)q_1(M_1-R_1) &=& 1\ ,\nonumber\\
  (M_1-R_1)(1+(m-2)q_1)+(M_D-R_D)q_1 &=& 0\ ,\nonumber
\eea

\noindent
where

\bea
d\mu(z)&\propto&  \exp(-\frac{|z|^2}{2M_1})
{\cal Z}(z)^m dz d\overline{z} ,\\
\langle f(C) \rangle_z &=& \int \dccc\  \nonumber
  \exp(- \beta |C|^{2\nu} -(M_D-M_1)|C|^2  - 2\Re(zC))  f(C)\ , \\
d\rho(h) &\propto&  exp(-\frac{|h|^2}{2R_1})\cosh^m(h_R)\cosh^m(h_I)
dh\ d\overline{h}\ , \\
\ \  \langle \tau \rangle_h &=& \tanh(h_R)+i \tanh(h_I)\ . \nonumber
\eea

We have not studied in detail the solutions of these equations, but
from the previous experience we conjecture that there is a transition
very similar to the Derrida model, and that such a transition
corresponds to a first step of replica symmetry breaking. We expect
the free energy lower bound we have obtained from the annealed
approximation to be very good.

%%%%%%%%%%%%%%%%%%%%%%%%%%%%%%%%%%%%%%%%%%%%%%%%%%%%%%%%%%%%%%%%%%%%%%%%%%
\section{A Discussion of the Phase Diagram\protect\label{S_ALL}}

If our initial conjecture about our effective theory and Hartree-Fock
resummation is correct we have solved the model in the high $T$ phase
(with two different approaches). That does not mean we have acquired a
large deal of information in the low $T$ phase.  Indeed the formulae
we have found cannot be valid at all temperatures since (analogously
to what happens in the GBD approximation) it leads to negative
entropies at low temperatures, and the entropy diverges logaritmically at
zero temperature.

\begin{figure}
  \epsfxsize=400pt\epsffile{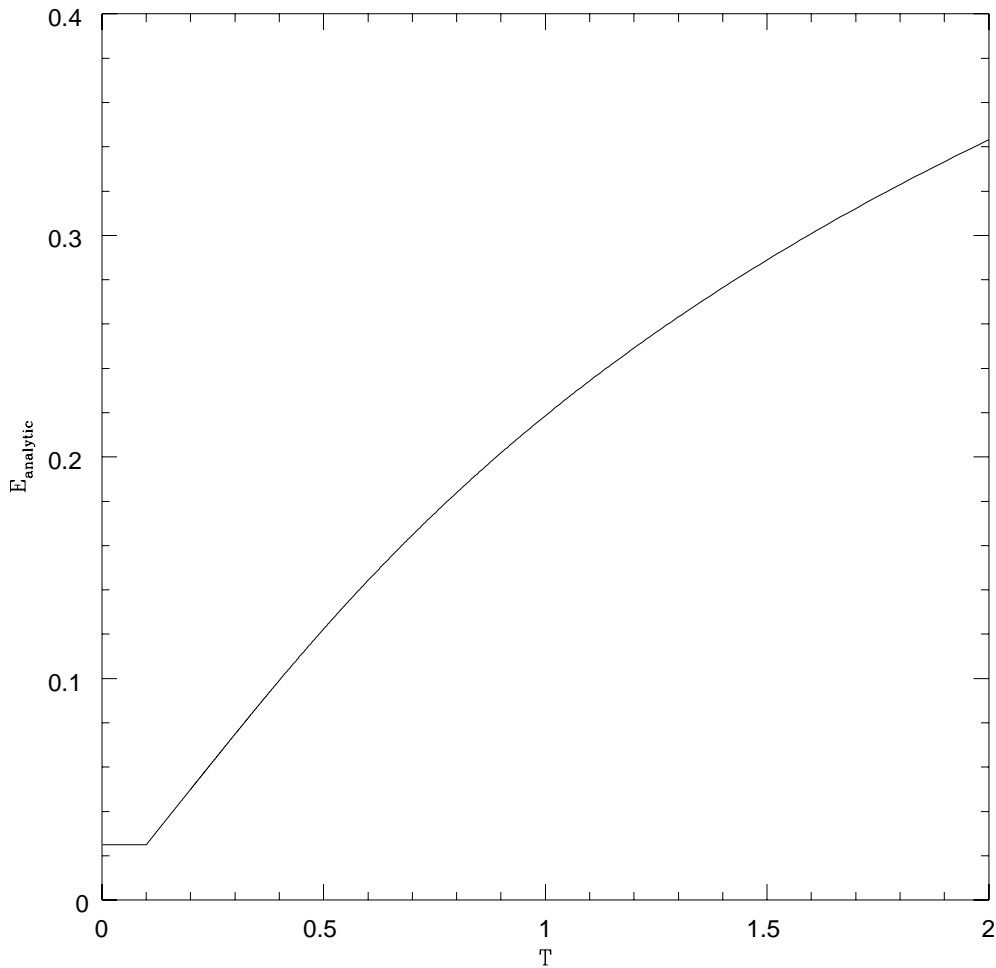}
  \caption[a]{\protect\label{F_EANA}
  The analytic result for the energy as a function of $T$.
  }
\end{figure}

We plot in fig. (\ref{F_EANA}) our result for the energy as a function
of $T$.  In our solution the energy goes to zero only at $T=0$. In an
approximation of the GBD type the entropy becomes zero at a non-zero
$T_G$, about $.1$, and the $T=0$ energy does not change in the cold
phase, and remains fixed to its value at $T_G$ and different from zero
(i.e. about $.025$). It is clear that we have to expect that the
high-temperature approximation breaks down before $T$ is lowered to
the point where the entropy is zero. More precisely it should break in
the region where the free energy is still negative, since the exact
result is that the free energy is zero at $T=0$ (at least for prime
values of $N$ and quite likely for all $N$).

\begin{figure}
  \epsfxsize=400pt\epsffile{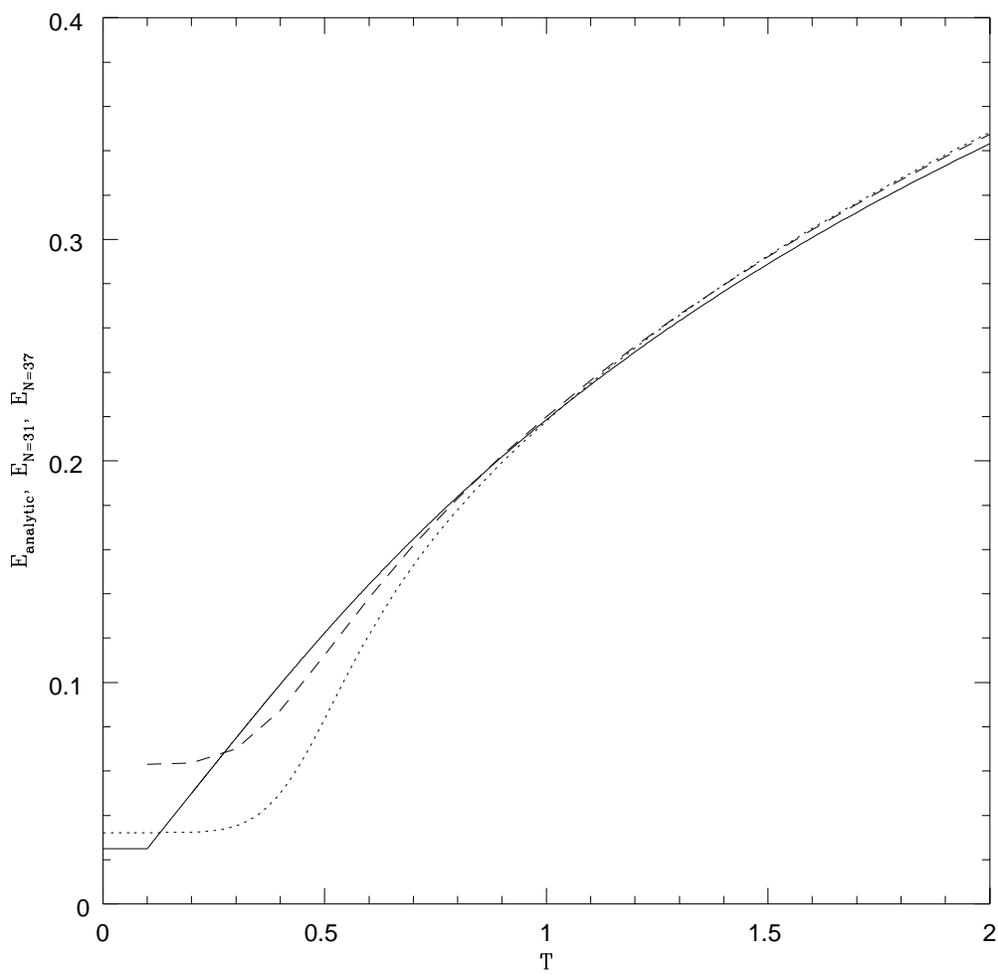}
  \caption[a]{\protect\label{F_EANAEXA}
  Comparison of the analytic results and the small $N$ exact
  solutions. Continuous curve for analytic solution, dots for $N=31$
  and dashes for $N=37$.
  }
\end{figure}

The comparison of these analytic results with the exact computations
is very interesting, and we show it in fig. (\ref{F_EANAEXA}).  In the
whole high temperature region where the energy varies from $1.0$ to
$0.2$ the agreement is very good, strongly supporting the correctness
of our solution in this temperature range. There is a disagreement in
the region where the energy becomes smaller and $T\to 0$.

The temperature where the free energy
becomes zero is about $T_F=.30$ (where the internal energy is about
$.074$).

At such low $T$ values the probability of finding the system in an
excited state, typically a single spin-flip of the ground state, is
negligible, since we know that the energy gap is at least of order
$3$. Let us draw a few possible, plausible scenarios:

\begin{itemize}

\item
The high-temperature approximation is valid down to a temperature very
close to $T_F$. At this temperature there is a first order phase
transition to a state with practically zero energy density. In this
case the discontinuity in the energy will be close to $.074$, and the
discontinuity in entropy close to $.25$. This is the possibility that
is favored from our evidence.

\item
The high-temperature approximation breaks down at a temperature higher
that $T_F$, i.e. about $T=.5$. In this case the transition could very
well be of the second order from the thermodynamic point of view. We
do not have any evidence for this possibility, but we cannot exclude it.

\item
It is also possible that when $N\to \infty$ for some values of $N$ the
energy density remains different from zero. If that is what is
happening the analytic results obtained by using replica theory could
be exact at all $T$ values, even in the low $T$ region for these
values of $N$. In other words we suggest the possibility that two
different thermodynamic limit can be obtained if we send $N \to\infty$
along different sequences. This would be a rather strange phenomenon
(which can happen only due to the infinite range nature of the
forces), however the non-unicity of the thermodynamic limit is present
in a related spin glass model~\cite{MAPARI2}.

\end{itemize}

{}From our results we are not able to discriminate in a definitive way
between these possibilities.

%%%%%%%%%%%%%%%%%%%%%%%%%%%%%%%%%%%%%%%%%%%%%%%%%%%%%%%%%%%%%%%%%%%%%%%%%%
\section{Conclusions\protect\label{S_CON}}

Let us summarize.  We have succeeded in obtaining a large body of
information about a deterministic system by using replica symmetry
theory. We have defined a deterministic, quite complex model, and in
first we have studied a simple approximation. We have shown that it is
easy to reproduce such simple approximation by using replica theory.
We have resummed the high temperature expansion of the model, and we
have shown that indeed replica theory allows to solve the model in the
whole high $T$ region. We have found indications about the nature of
the transition regime, but we have not been able to describe in detail
the transition point and the low $T$ phase.

In order to get the bulk of our analytical results we have written a
disordered model, where we have substituted the Fourier transform with
a generic unitary transformation (after some thinning of degrees of
freedom). The two models coincide at high temperature, but they do
(very probably) differ at low temperature. The deterministic model has
(very probably) zero energy density at zero temperature, while the
second one has a ground state energy density equal to $.025$.  Still,
we have to note that if for generic values of $N$ (non good primes,
where we know we get a zero energy density) the deterministic system
would admit a ground state with energy equal to $.025 N$, we could
appreciate the effect only for $N$ very large, of the order of $200$,
while we have been able to solve the model only up to $N=38$.  We
cannot exclude that in the deterministic model the energy density is
indeed non-zero for generic values of $N$, or even that for different
choices of $N$ (of non zero measure) one could get different
behaviors.

We believe that the use of replica field theory for studying systems
without quenched noise is a very promising tool, which will be able to
lead to precise results both in the high and in the low temperature
phase. Systems without built-in disorder can have a complex landscape,
and one can use replica theory to understand it.

%%%%%%%%%%%%%%%%%%%%%%%%%%%%%%%%%%%%%%%%%%%%%%%%%%%%%%%%%%%%%%%%%%%%%%%%%%
\section*{Acknowledgements}

We are more than happy to acknowledge very useful discussions and a
continuous fruitful collaboration on subjects related to the one of
this paper with Leticia Cugliandolo, Silvio Franz, Jorge Kurchan, Marc
Mezard, Gabriele Migliorini and Miguel Virasoro. In particular we are
grateful to Marc Mezard for discussing with us his results prior to
publication.

\end{document}